\documentclass[journal]{vgtc}                     

\DeclareGraphicsExtensions{.png,PNG,.pdf,.PDF}

\onlineid{1377}

\vgtccategory{Research}

\vgtcpapertype{algorithm/technique}

\title{Cluster-Aware Grid Layout}

\author{%
  Yuxing Zhou,
  Weikai Yang,
  Jiashu Chen,
  Changjian Chen,
  Zhiyang Shen,
  Xiaonan Luo,
  Lingyun Yu,
  and Shixia Liu
}
\authorfooter{
  \item
  	Y.~Zhou, W.~Yang, J.~Chen, Z.~Shen, and S.~Liu are with the School of Software, BNRist, Tsinghua University. Y.~Zhou and W.~Yang are joint first authors. S.~Liu is the corresponding author.
  	E-mail: \{\{yx-zhou19, yangwk21, cjs22, shenzhiy21\}@mails., shixia@\}tsinghua.edu.cn.
   \item
       C.~Chen is with Kuaishou Technology.
       E-mail: chenchangjian@kuaishou.com
  \item
  	X.~Luo is with Guilin University of Electronic Technology.
  	E-mail: luoxn@guet.edu.cn.
  \item L.~Yu is with Xi'an Jiaotong-Liverpool University.
  	E-mail: Lingyun.Yu@xjtlu.edu.cn.
}
\abstract{
Grid visualizations are widely used in many applications to visually explain a set of data and their proximity relationships. 
However, existing layout methods face difficulties when dealing with the inherent cluster structures within the data. 
To address this issue, we propose a cluster-aware grid layout method that aims to better preserve cluster structures by simultaneously considering proximity, compactness, and convexity in the optimization process.
Our method utilizes a hybrid optimization strategy that consists of two phases.
The global phase aims to balance proximity and compactness within each cluster, while the local phase ensures the convexity of cluster shapes. 
We evaluate the proposed grid layout method through a series of quantitative experiments and two use cases, demonstrating its effectiveness in preserving cluster structures and facilitating analysis tasks.
}


\teaser{
  \centering
  \includegraphics[width=\linewidth]{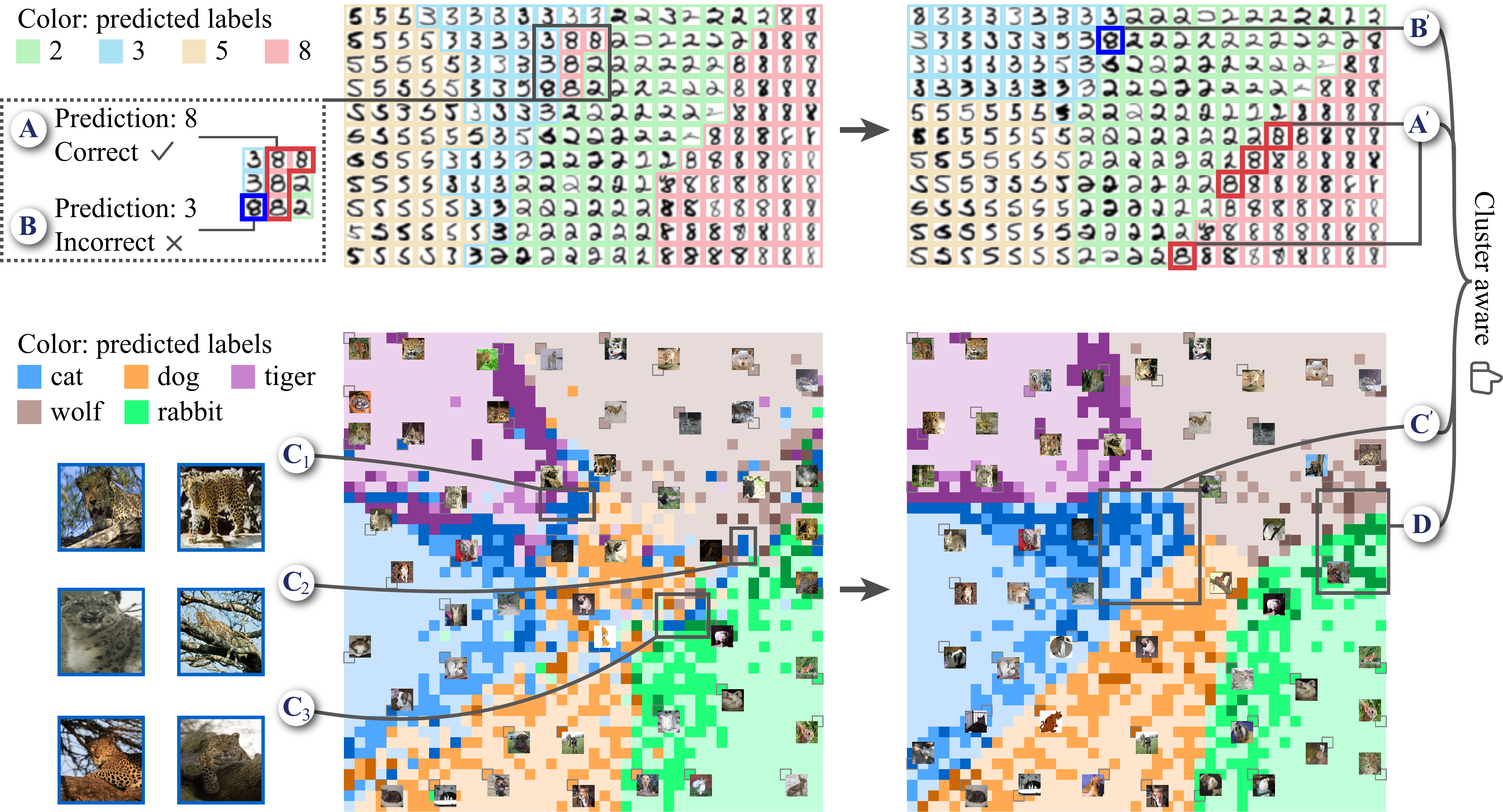}
  \put(-308,159){\rm{(a) Baseline}}
  \put(-136,159){\rm{(b) Our method}}
    \put(-308,-10){\rm{(c) Baseline}}
  \put(-136,-10){\rm{(d) Our method}}
  \vspace{-1mm}
  \caption{
  \weikai{Top: compared with the baseline (a) that only optimizes proximity, our method (b) places the samples that are predicted as ``8'' ($\text{A}$) in the cluster of ``8.''
  Bottom: compared with the baseline (c), our method (d) groups the dark blue cells C${}_1$-C${}_3$ in region C'.}
  }
  \vspace{-1mm}
  \label{fig:teaser}
}

\graphicspath{{figs/}{figures/}{pictures/}{images/}{./}} 

\usepackage{multirow}
\usepackage{color}
\usepackage{lipsum}                    
\usepackage{amssymb,amsmath}
\usepackage{wrapfig}
\usepackage{booktabs}
\usepackage{makecell}
\usepackage{paralist}
\usepackage{bm}
\def \etal {{\emph{et al}.\thinspace}}
\def \eg {{\emph{e.g}.\thinspace}}
\def \ie {{\emph{i.e}.\thinspace}}
\newcommand{\HP}[1]{\textbf{H{#1}}}
\newcommand{\E}[1]{E{#1}}
\newcommand{\shixia}[1]{\textcolor{black}{#1}}
\newcommand{\weikai}[1]{\textcolor{black}{#1}}
\newcommand{\glyph}[2][1.3]{$\vcenter{\hbox{\includegraphics[height=#1\fontcharht\font`\B]{fig/glyph/#2.pdf}}}$}

\def \lightbrownbox{{\hbox{\pdfliteral{0.9  0.86  0.86 rg}\vrule height0.2cm width0.2cm depth0cm\pdfliteral{0 g}}}}
\def \lightpurplebox{{\hbox{\pdfliteral{0.89  0.83  0.91 rg}\vrule height0.2cm width0.2cm depth0cm\pdfliteral{0 g}}}}
\usepackage{mathptmx}                  

\begin{document}


\firstsection{Introduction}

\maketitle
\fontsize{9}{9} 
\firstsection{Introduction}
\maketitle

Grid visualizations are widely used to {visually analyze} data collections due to their high space efficiency~\cite{frey2022optimizing}.
Over two hundred {CVPR 2022} papers utilize grid-like visualizations to compare and analyze model outputs.
{With} these visualizations, {computer vision researchers} hope to perceive the samples in one cluster (\eg, the samples with the same predictions) as a whole, {which makes it easier to diagnose potential causes of low-performance models}.
If such cluster structures are not well perceived, accurate analysis {and diagnosis} will be hindered.
{For example, in the baseline method that only considers proximity (\cref{fig:teaser}(a)), the samples of ``8'' in $\text{A}$ and $\text{B}$ are placed far away from the cluster of ``8.''
This arrangement may lead users to draw the wrong conclusion that the model predicts these samples to be closely related to ``3.''
However, the samples in $\text{A}$ are similar to other samples of ``8'' and are predicted as ``8.''
By preserving cluster structures, these samples are merged into the cluster of ``8,'' which reduces false inferences.
In addition, the sample that is misclassified as ``3'' ($\text{B}$) remains in the cluster of ``3'' (\cref{fig:teaser}$\text{B}'$), which can be identified.}


Several grid layout methods have been developed to improve readability by preserving the proximity relationships between data samples~\cite{fried2015isomatch,quadrianto2008kernelized,strong2014self}.
Despite their benefits, these methods struggle to maintain cluster structures within the data.
{For example, the samples in \cref{fig:teaser}A are placed far away from their corresponding cluster.}
According to the Gestalt principles of perceptual grouping, preserving cluster structures requires not only the preservation of {\emph{proximity}} relationships but also the {\emph{compactness}} and {\emph{convexity}} of each cluster shape~\cite{todorovic2008gestalt,rottmann2022mosaicsets,kanizsa1976convexity}.
To develop a layout method that {considers} {all three principles simultaneously, it is crucial to quantify them}.
Proximity {is usually measured by the similarity preservation between samples and} has been well studied by existing grid layout methods.
The compactness is usually measured by the deviation of the grid positions from their corresponding cluster centers~\cite{rottmann2022mosaicsets}. 
However, there is currently no widely accepted measure for quantifying shape convexity that aligns well with people's perception.
To address this issue, we conducted a user study with 54 participants to evaluate which convexity measures are more consistent with people's perception.
We found that although no single measure matched all participants' perception, two representative measures were preferred by two distinct groups of participants.
However, these two measures conflict with each other to some extent.
{This requires our method to support different convexity measures to meet diverse user preferences}.\looseness=-1

After quantifying proximity, compactness, and convexity, we develop a {cluster-aware} grid layout method that balances the three measures.
{However, achieving this balance is challenging, especially when attempting to consider all three measures simultaneously during the layout process.}
Upon analyzing these measures, we discovered that proximity and compactness are {affected by all grid cells}, while convexity is sensitive to {boundary cells} between different clusters.
Based on {this finding}, our layout method {employs} a global–local strategy to simplify the optimization process.
{Accordingly, the layout method consists of two phases: global assignment and local adjustment.}
{The global assignment phase aims to generate a layout that balances proximity and compactness.
This is formulated} as a multi-task linear assignment problem and solved by an accelerated Jonker-Volgenant algorithm~\cite{chen2020oodanalyzer}.
{The local adjustment phase attempts} to swap {boundary cells} between different clusters to improve convexity {without apparently compromising the proximity and compactness achieved in the global assignment}.
{Quantitative experiments} demonstrate that our layout method achieves {experimentally optimal} balances among proximity, compactness, and convexity.
{We also present two use cases to exemplify the usage of our method.}\looseness=-1

The main contributions of our work include:
\begin{compactitem}
\item\noindent {study results on which convexity measures are more} consistent with human perception.
\item\noindent a grid layout method that {achieves experimentally} optimal balance among proximity, compactness, and convexity.
\item\noindent an open-source implementation of the proposed grid layout method that enables easy plug-in of different convexity measures, {which is available at \href{https://github.com/thu-vis/Cluster-Aware-Grid-Layout}{https://github.com/thu-vis/Cluster-Aware-Grid-Layout}}.
\end{compactitem}

\section{Related Work}
\label{sec:related-work}

\subsection{Convexity Measures}
\label{subsec:convexity}
{Mathematically, a} shape is convex if it completely contains the line segment connecting any two points within the shape~\cite{rosin2007probabilistic}.
{Based on this definition, researchers have developed various convexity measures, which can be classified into two categories~\cite{rahtu2006new}}: area-based measures and boundary-based measures.

Area-based measures {rely on the area of the shape to determine their scores}.
{A common measure for evaluating the convexity of a shape is the area ratio, which computes the ratio of its actual area to {that} of its convex hull}~\cite{efrat2014mapsets,sonka2014image}.
This measure was extended by using the largest convex polygon contained in the shape (convex skull)~\cite{zunic2004new} or considering the ratio between the area of the convex skull and the area of the convex hull~\cite{bozeman2013convexity}.
However, these measures are sensitive to {long and} thin protrusions \glyph{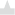} or intrusions \glyph{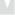} because such protrusions/intrusions will {largely} affect the shape of its convex hull/skull.
{To address this issue,} Rosin and Mumford~\cite{rosin2006symmetric} improved this measure {by calculating the discrepancy} between the area of the shape and its maximally overlapping convex shape.
{In addition to considering the area discrepancy, researchers have proposed several measures based} on probability.
For example, Held and Abe~\cite{held1994approximate} estimated the degree of convexity by computing the probability that the shape contains line segments connecting two randomly sampled points inside the shape.
{Rahtu~\etal\cite{rahtu2006new} proposed a faster computation method by verifying if the shape contains a specific point on the segment (e.g., the midpoint) instead of examining the entire segment}.
{Recently, \v{Z}uni\'{c} and Rosin~\cite{vzunic2019measuring} proposed a parametric measure based on the similarity between the shape and its convex hull.
A larger parameter of the measure leads to a stronger penalty for the area discrepancy near the boundary.
}

Boundary-based measures {evaluate the convexity of a shape by analyzing the geometric properties of its boundary}, such as perimeter and tangents.
{A basic measure} {is the perimeter ratio, which computes} the ratio of the $L^2$ perimeter of the convex hull to that of the shape~\cite{peura1997efficiency}.
\v{Z}uni\'{c} and Rosin~\cite{zunic2004new} further proposed to use the minimum ratio of {the} $L^2$ perimeter of its bounding rectangles to the $L^1$ perimeter {of the shape} over all the rotations, which is more sensitive to changes in the boundary of a concave region.
Another boundary-based measure is proposed based on the fact that a convex shape always lies entirely on one side of its tangent~\cite{do2016differential}.
The convexity measure is then {measured by} the average ratio over all the dominant parts cut by the tangents.

\begin{figure}[!tb]
\centering
\includegraphics[width=\linewidth]{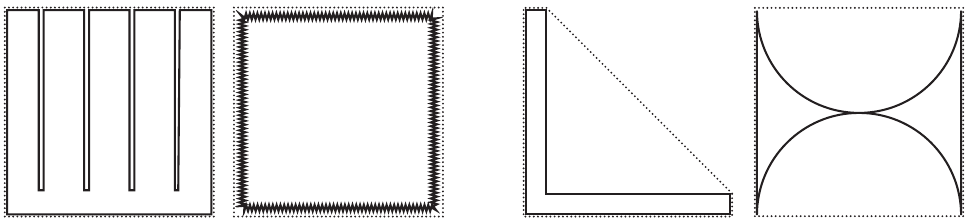}
\put(-249,-13){(a) 0.95, 0.50}
\put(-188,-13){(b) 0.96, 0.31}
\put(-116,-13){(c) 0.32, 0.87}
\put(-54,-13){(d) 0.21, 0.78}
\caption{The comparison between area ratios (the first value) and perimeter ratios (the second value). In (a) and (b), the boundary length changes considerably while the area changes slightly. The low perimeter ratios indicate that the perimeter ratio is more sensitive to changes in the boundary length. In contrast, (c) and (d) show considerable changes in their area and slight changes in boundary length.
The low area ratios indicate that the area ratio is more sensitive to changes in the area.}
\vspace{-4mm}
\label{fig:convexity}
\end{figure}

Generally, area-based measures are sensitive to changes in the area of a shape, while boundary-based measures are sensitive to changes in the boundary of a shape.
\cref{fig:convexity} compares the area ratio (area-based) and perimeter ratio (boundary-based) in four examples.
The perimeter ratios of the two examples with considerable changes in their boundary length ((a) and (b)) are much lower.
While the area ratios of the two examples with considerable changes in their area ((c) and (d)) are much lower.
 The selection of convexity measures will depend on the specific analysis tasks.
{If users want to detect large concavities in the area, they usually choose area-based measures, while boundary-based measures are preferred for detecting irregular boundaries. 
To determine which measures are better aligned with human perception in a grid visualization, we conducted a user study to identify the most appropriate measures. These measures are given priority in our layout method.}\looseness=-1

\subsection{Grid Visualizations}
\label{subsec:layout}
Initial efforts {on grid visualizations randomly assign data samples to the grid cells}~\cite{matejka2018dream}.
Despite its simplicity, {this method has proven useful for visually analyzing various data, including} images~\cite{song2016gazedx,Liu2017,huang2022conceptexplainer,oppermann2021vizsnippets,chen2022towards}, textual data~\cite{felix2016texttile,sevastjanova2022visual}, video data~\cite{chan2019motion,keefe2009interactive}, relational data~\cite{muelder2008rapid,yuan2022visual,major2018graphicle}, geometric data~\cite{matejka2018dream,choi2021dxplorer}, and geospatial data~\cite{zeng2020revisiting,yang2022epimob}. 
Subsequently, many grid layout methods have been developed {to facilitate the analysis of similar samples by preserving pairwise distances between them. 
These methods fall} into two categories~\cite{halnaut2022vrgrid}: direct mapping methods and projection-based methods.

{The methods in the first category} directly map high-dimensional samples onto {a two-dimensional grid}~\cite{barthel2022improved,tu2022phrasemap}.
Quadrianto~\etal\cite{quadrianto2008kernelized} {proposed a method to maximize} the {correlation} between the {pairwise} distances in the high-dimensional space and the {pairwise} distances in the grid layout.
{Another method, the} self-sorting map~\cite{strong2014self}, randomly assigns samples to grid cells and then iteratively swaps them to improve {the similarity between neighboring samples}.
Barthel and Hezel~\cite{barthel2019visually} further improved the {proximity preservation in} the self-sorting map {by utilizing an adaptive method to calculate the neighborhood representation of each sample}.
However, these two methods use a brute-force search to find {the best swap}, which is time-consuming.
{To boost efficiency, Barthel~\etal\cite{barthel2022improved} identified the best swap by solving a small-scale linear assignment problem locally}.
{Other methods aim to optimize additional measures}, such as compactness~\cite{meulemans2016small,rottmann2022mosaicsets} and aesthetic criteria~\cite{yoghourdjian2015high,pan2019content,song2021balance}.
{To handle} a large number of samples,
Frey~\cite{frey2022optimizing} {generated a} hierarchical grid layout by simultaneously optimizing the similarity between neighbors and the homogeneity within each node in the hierarchy.
{
There are some treemap-based methods that visualize samples in a hierarchical grid format~\cite{bederson2001photomesa,bertucci2023dendromap}. However, these methods do not consider the proximity between samples within each cluster.
}

Projection-based methods first utilize a dimensional reduction technique to project the samples onto a 2D space without the grid constraints.
Then the final layout is generated by moving samples from the projected positions to the grid cells~\cite{eppstein2015improved,fried2015isomatch,hilasaca2019overlap,halnaut2022vrgrid}.
For example, IsoMatch~\cite{fried2015isomatch} uses IsoMap to project images onto a 2D space.
After building the bipartite graph between the projected positions and grid cells, the grid layout is obtained by finding the bipartite matching that minimizes the total distance moved.
Since it is time-consuming to solve the bipartite matching problem, later studies {attempt} to accelerate the process of assigning projected samples into cells.
Chen~\etal\cite{chen2020oodanalyzer} developed a $k$NN-based bipartite graph matching, which speeds up the algorithm by reducing the number of candidate grid cells for each sample.
{Additionally,} a grid layout can be generated by removing the overlap between projected samples and then aligning them. 
DGrid~\cite{hilasaca2019overlap} recursively bisects the projected samples until each partition contains exactly one sample.
The partition result is then aligned with a regular grid layout.
CorrelatedMultiples~\cite{liu2018correlatedmultiples} uses a force-directed graph layout method to remove the overlap between samples and then aligns the samples horizontally and vertically to form a grid layout.
{This technique is also used to lay out a clustered graph in a grid format~\cite{itoh2009hybrid}.}
However, when the samples are unevenly projected on a 2D space, these alignment techniques can result in large movements from projected positions to grid cells.
To address this issue, VRGids~\cite{halnaut2022vrgrid} employs the Voronoi relaxation to scatter the projected samples evenly on a bounded 2D space before assigning them to cells, which reduces the total movements.

Although the aforementioned methods have been proven effective in preserving the proximity relationships between data samples,
they do not explicitly consider the cluster structures within the data.
{Consequently}, they face difficulties when tasked with preserving such structures. 
Our layout method starts from a grid layout generated by {any of} these methods and then {enhances the compactness and convexity of each cluster shape to preserve cluster structures}.
When {handling} a large number of samples, a flat grid layout {encounters} the scalability issue in displaying all the samples {clearly in one view}.
Our method can either utilize the existing hierarchy in the input layout or build the hierarchy using sampling techniques~\cite{chen2020oodanalyzer}.
At each level of the hierarchy, {our method enhances compactness and convexity to preserve the cluster structures}.\looseness=-1

\begin{figure}[!tb]
\centering
\includegraphics[width=0.95\linewidth]{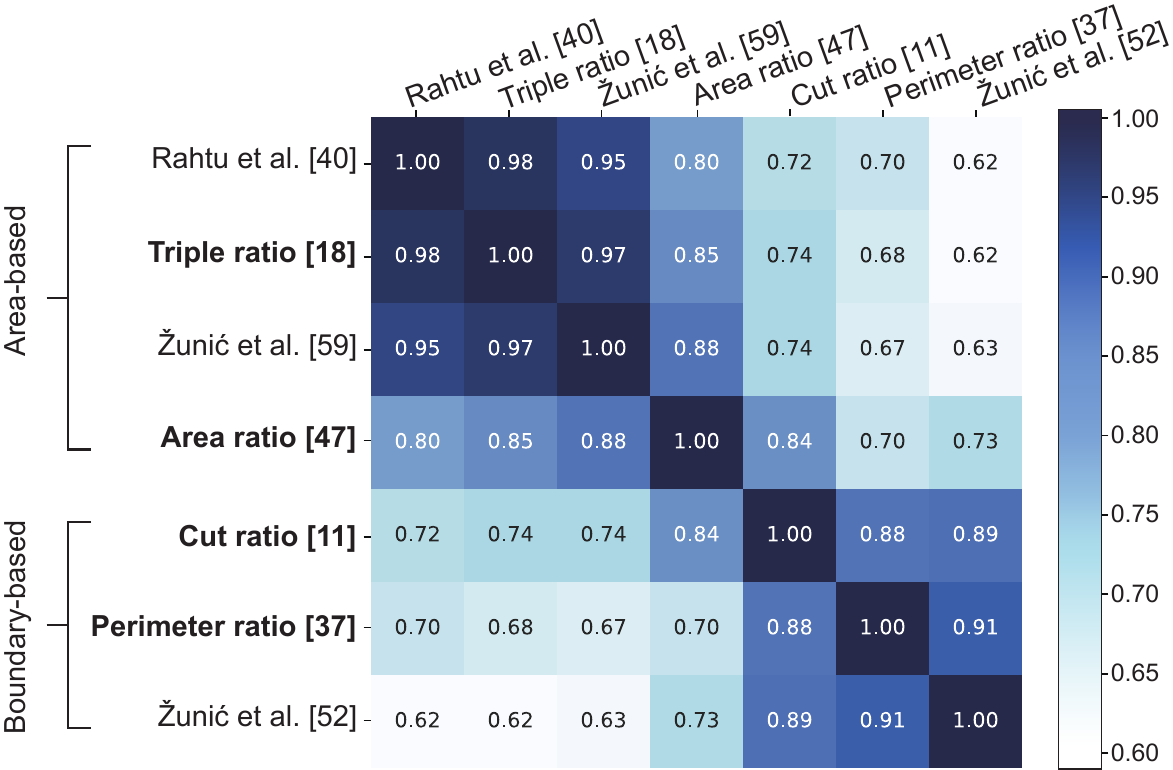}
\caption{Pearson correlations between seven convexity measures. {The block-diagonal pattern highlights the presence of two different clusters.}}
\label{fig:correlation}
\vspace{-4mm}
\end{figure}

\section{User Study on Convexity}
\label{sec:requirement}

{
{The user study has two goals:} 
first, to determine which measures {closely align} with human perception, and second, to explore whether variations in grid size and cluster number {influence} {people's} perception of convexity.
To achieve these goals}, three hypotheses have been formulated {to} guide the design of our user study:\\
\HP1: There exists a convexity measure that {aligns} with the perception of most people.\\
\HP2: The grid size {{influences} people's} perception of convexity.\\
\HP3: The cluster number {{influences} people's} perception of convexity.\\

\subsection{User Study Design}
\label{subsec:userstudy}
\noindent\textbf{Selection of convexity measures.}
For the 10 convexity measures introduced in \cref{subsec:convexity}, we excluded 3 measures due to their high time complexity:
two of them require computing convex skulls with a time complexity of $O(N^7)$~\cite{zunic2004new,bozeman2013convexity}, and one requires finding maximally overlapping convex shapes with a time complexity of $O(2^N)$~\cite{rosin2006symmetric} ($N$ is the number of vertices).
\cref{fig:correlation} shows the Pearson {correlations} between the implemented convexity measures calculated on 9,689 different shapes. 
{Further details of this experiment are summarized} in {supplemental} material.
{The results in \cref{fig:correlation} indicate the seven measures are roughly classified into two clusters 
that correspond to the area-based measures and boundary-based measures, respectively}.
The correlations between four area-based measures are strong, and so are the correlations between three boundary-based measures.
Due to the high correlations, we selected the representative measures rather than using all of them in the user study, {thus} making it {easier} for participants to compare multiple grid visualizations optimized for different measures.

Among the four area-based measures, \emph{area ratio}~\cite{sonka2014image} is selected first because it has a relatively weak correlation ($r<0.9$) with the other three measures {and cannot be represented by them}.
{Of the remaining three highly-correlated area-based measures,} we selected \emph{triple ratio}~\cite{held1994approximate} since it has stronger correlations with the other two measures (0.98 and 0.97).
In addition, \emph{triple ratio} is more efficient to be computed than \v{Z}uni\'{c}'s method~\cite{vzunic2019measuring}, and also provides more precise measurements than Rahtu's method~\cite{rahtu2006new}.
Among the three boundary-based measures, \emph{perimeter ratio}~\cite{peura1997efficiency} and \v{Z}uni\'{c}'s method~\cite{zunic2004new} {are both based on the perimeter and are highly correlated (0.91)}.
We selected \emph{perimeter ratio} because its calculation is more efficient ($O(N\log N)$) than the latter ($O(N^2)$).
{We also chose to include \emph{cut ratio}}~\cite{do2016differential} because it has a relatively weaker correlation ($r<0.9$) with the other two measures.
The definitions of the four selected measures are provided below:

\begin{wrapfigure}{l}{0.066\textwidth}
\vspace{-12pt}
\includegraphics[width=0.066\textwidth]{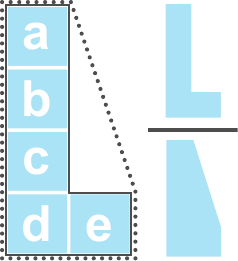}
\end{wrapfigure}
\noindent{\emph{Area ratio} ({A})} 
is defined as the ratio of the area of shape to the area of its convex hull.
In this example, the length of each side of the squares is 1.
Thus, the area of the shape is 5, and the area of the convex hull is $5+1.5=6.5$.
The convexity score is calculated as $5/6.5\approx 0.769$.

\begin{wrapfigure}[5]{l}{0.068\textwidth}
 \vspace{-12pt}
\includegraphics[width=0.068\textwidth]{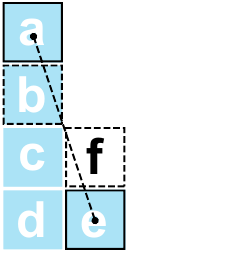}
\end{wrapfigure}
\noindent{\emph{Triple ratio} ({T})}
is defined as the probability that, for a collinear triple $(X,Y,Z)$, if both cells $X$ and $Z$ are inside the shape, then cell $Y$ is also inside the shape.
In this example, there are 8 collinear triples of which both endpoints are {located} inside the shape (\eg, $(a,b,c)$ and $(a,b,e)$).
However, the interior cell of 2 triples ($(a,f,e)$ and $(b,f,e)$) {lies} outside the shape.
Thus, the convexity score is calculated as $(8-2)/8=0.75$.

\begin{wrapfigure}[5]{l}{0.066\textwidth}
 \vspace{-12pt}
\includegraphics[width=0.066\textwidth]{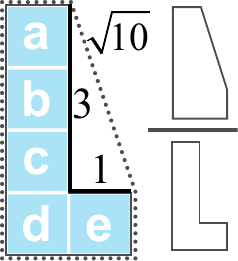}
\end{wrapfigure}
\noindent{\emph{Perimeter ratio} ({P})}
is defined as the ratio of the perimeter of the convex hull to the perimeter of the shape.
In this example, the perimeter of the convex hull is $8+\sqrt{10}$, and the perimeter of the shape is $8+4=12$.
The convexity score is calculated as $(8+\sqrt{10})/12\approx 0.857$.

\begin{wrapfigure}[5]{l}{0.066\textwidth}
\vspace{-14pt}
\includegraphics[width=0.066\textwidth]{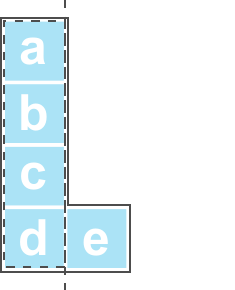}
\end{wrapfigure}
\noindent{\emph{Cut ratio} ({C})}
is proposed based on the {property that for a convex shape, all tangents to its boundary will not intersect its interior}.
It calculates, for each edge in the boundary, the ratio of the part cut by the tangent.
The convexity measure is then defined as the {average ratio over all the edges.}
In this example, the convexity score is calculated as $(1\times8+0.8\times3+0.4\times1)/12=0.9$.

\noindent\textbf{Participants}.
We recruited 54 participants (44 male and 10 female) in the experiment, including faculty members, {researchers, developers,} and graduate students with {industry/research} experience in visualization, computer graphics, computer vision, or mathematics.
{
They have used grid layouts to explore data in their research/development.
The participants come from 3 countries and 9 institutions.
The diversity in expertise, experience, and affiliations ensures that our recruited participants are representative.}
{Among the 54 participants, all have knowledge of convexity, none reported color blindness or color weakness, and 41 of the participants are very familiar with grid visualizations}.
Upon completion, each participant was rewarded a \$20 gift card.

\noindent\textbf{Study procedure}.
At the beginning of the study, participants were presented with a tutorial video that introduced the definition of convex polygons and the user interface of the study system.
{After watching the video, the participants began a practice session with six trials.
{The answers and corresponding explanations were displayed after participants submitted their results.}
After completing the practice {session} and indicating their full understanding of the concept of convex polygons and the study interface, the participants proceeded to the formal study with 36 trials, in which answers were no longer displayed.}
Participants were instructed that they could take a brief break after completing every nine trials.
{Following the completion of all trials, they were asked to fill out a questionnaire that included personal information and a question asking them to {explain how they} compare the convexity of different grid visualizations}.
The entire process took approximately 40 minutes.
{The study received approval from the University Ethics Committee.}

\noindent\textbf{{Trials and stimuli}}.
To validate \HP1, in each formal trial, the participants were asked to rank four different grid visualizations, each optimized for one of the four {selected} convexity measures.
This enabled the ranking results to reflect their preference for the convexity measures.
{Before the formal study, we provided six practice trials} to ensure that the participants correctly understood the concept of convex polygons.
The stimuli in each practice trial {consisted of four grid visualizations arranged in descending order} on all four convexity measures.

\noindent\textbf{{Conditions and design}}.
To {validate \HP2 and \HP3, we manipulated two variables, the \emph{grid size} and the \emph{cluster number}, to control for their effects.}
In {practical applications, the maximum grid size is typically restricted to 40x40 to ensure that each cell has enough space to display data clearly}.
Therefore, we chose three different grid sizes: 20x20, 30x30, and 40x40 in the experiment.
We selected three cluster numbers of 3, 5, and 10 because {analyzing a large number of clusters simultaneously can be challenging to the participants.
}
 To generate trials for each condition, we used four datasets, Animal~\cite{dataset-animals}, MNIST~\cite{dataset-mnist}, CIFAR10~\cite{dataset-cifar10}, and USPS~\cite{dataset-usps}, each of which contains 10 clusters.
{A full-factorial within-subjects design was used to evaluate the effects of the grid size and cluster number}.
{As a result, for each participant, a total of 36 trials (3 grid sizes $\times$ 3 cluster numbers $\times$ 4 datasets)} were evaluated in the formal study.
{The orders for the grid sizes and cluster numbers were counterbalanced using a Latin Square Design.}

\subsection{Result Analysis}
\label{subsec:userstudyresult}
\noindent\HP1: There exists a convexity measure that {aligns} with the perception of most people {(partially confirmed)}.

\begin{figure}[!tb]
\centering
\includegraphics[width=\linewidth]{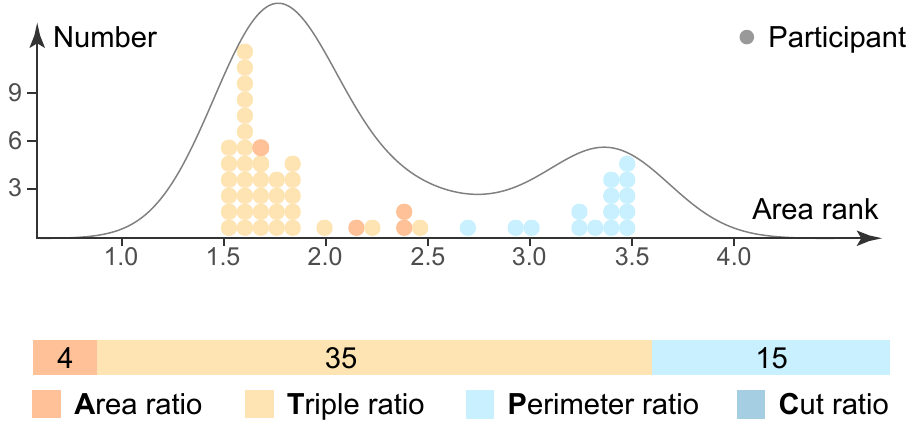}
\put(-220,34){(a) Distribution of area ranks of 54 participants.}
\put(-240,-10){(b) Distribution of the most preferred measures of 54 participants.}
\caption{Examine the distribution of area ranks and most preferred measures of 54 participants. {Among these participants, 39 preferred area-based measures (orange), while the other 15 preferred boundary-based measures (blue).}}
\vspace{-5mm}
\label{fig:mostpreferred}
\end{figure}

We analyzed the ranking results to determine if {there is a specific convexity measure that was preferred by the majority of the participants}.
We first processed the ranking result of each participant.
If a participant chooses $=$ between two measures, their ranks will be the average of the ranks.
For example, if a participant ranks the measures as $\textbf{A}>\textbf{T}>\textbf{P}=\textbf{C}$, their ranks will be 1, 2, 3.5, and 3.5.
{The ranking result of each participant is computed by averaging his/her ranks over all the trials.}
{Next, we conducted Friedman tests to compare the ranks of different measures on all the participants.
{However, the tests showed no significant differences between the measures.}
Upon further examination of the ranking results, we observed that there was a large variance in the rankings for \textbf{T} and \textbf{P}: some participants ranked \textbf{T} highest and ranked \textbf{P} lowest, while others ranked \textbf{P} highest and ranked \textbf{T} lowest. 
This diversity led to the large variances and made the differences not significant.}
In addition, we found a strong correlation between two boundary-based measures \textbf{P} and \textbf{C} (0.955).
This indicates that participants who preferred the boundary-based measure \textbf{P} also tended to prefer the boundary-based measure \textbf{C}.
Similarly, a strong correlation was found between the two area-based measures \textbf{A} and \textbf{T} (0.926).
{We obtained the \emph{boundary rank} for each participant by averaging the ranks of \textbf{P} and \textbf{C}, and the \emph{area rank} by averaging the ranks of \textbf{A} and \textbf{T}}.
{It is important to note that} the sum of the boundary rank and area rank always adds up to five.
Therefore, it is sufficient to analyze only one of them.
\cref{fig:mostpreferred}(a) shows the distribution of area rank.
There were two distinct groups of participants, where 39/54 (72.2\%) of them preferred the area-based measures (area rank$<2.5$), while the remaining (15/54, 27.8\%) preferred the boundary-based measures (area rank$>2.5$).

\begin{figure}[!b]
\centering
\includegraphics[width=\linewidth]{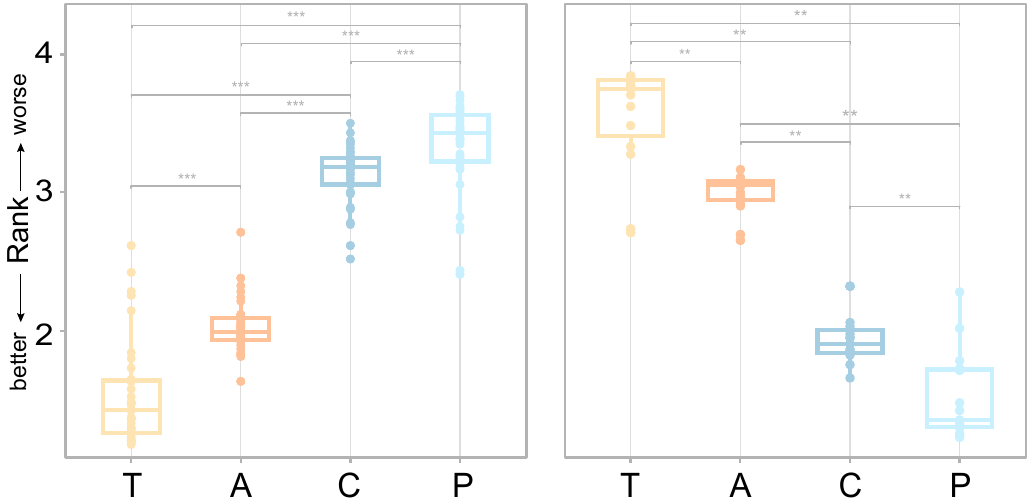}
\put(-228,127){\small$\chi^2(3)=98.64,n=39,p<0.001$}
\put(-105,127){\small$\chi^2(3)=43.88,n=15,p<0.001$}
\caption{Friedman tests and pairwise Wilcoxon signed-rank tests on four measures.
Left: the participants who preferred area-based measures {have a strict preference order of $\textbf{T}>\textbf{A}>\textbf{C}>\textbf{P}$}; Right: the participants who preferred boundary-based measures {have a strict preference order of $\textbf{P}>\textbf{C}>\textbf{A}>\textbf{T}$}.
Significance levels are denoted by asterisks: 
* indicates $p<0.05$, ** indicates $p<0.01$, and *** indicates $p<0.001$.}
\label{fig:friedman}
\end{figure}

Next, we investigated the measures that were most preferred by each participant.
Surprisingly, 35 of them ranked \textbf{T} highest, and 15 of them ranked \textbf{P} highest (\cref{fig:mostpreferred}(b)).
The result indicates that most of the participants who preferred area-based measures ranked \textbf{T} higher than \textbf{A} (35/39, 89.7\%), whereas all the participants who preferred boundary-based measures ranked \textbf{P} higher than \textbf{C} (15/15, 100.0\%).
We thus performed Friedman tests {again} to compare the ranks of different measures for the two groups of participants separately.
{The results showed that for participants who preferred area-based measures (\cref{fig:friedman} Left), the difference among the four measures was significant} ($\chi^2(3)=98.64,\ p<0.001$).
{The corresponding effect size was $0.8612$, which also indicated a great difference between different measures.}
The pairwise Wilcoxon signed-rank test results {further} indicated a strict preference order of $\textbf{T}>\textbf{A}>\textbf{C}>\textbf{P}$.
{For participants who preferred boundary-based measures (\cref{fig:friedman}, Right), the difference among the four measures was also significant} ($\chi^2(3)=43.88,\ p<0.001$), and the pairwise Wilcoxon test results showed a strict preference order of $\textbf{P}>\textbf{C}>\textbf{A}>\textbf{T}$. 
The corresponding effect size was $0.9751$, which again indicated a great difference between different measures.
{It is important to note that this order is} the exact reverse of the order found in participants who preferred area-based measures.
Based on these findings, we {concluded that} no single measure aligns perfectly with the perception of all participants.
However, measure \textbf{T} and measure \textbf{P} are the representative of area-based measures and boundary-based measures, respectively.
For those who preferred area-based measures, measure \textbf{T} aligns best with their perception, and for those who preferred boundary-based measures, measure \textbf{P} aligns best with their perception.
Therefore, \HP1 is partially confirmed.

\begin{figure}[!tb]
\centering
\includegraphics[width=\linewidth]{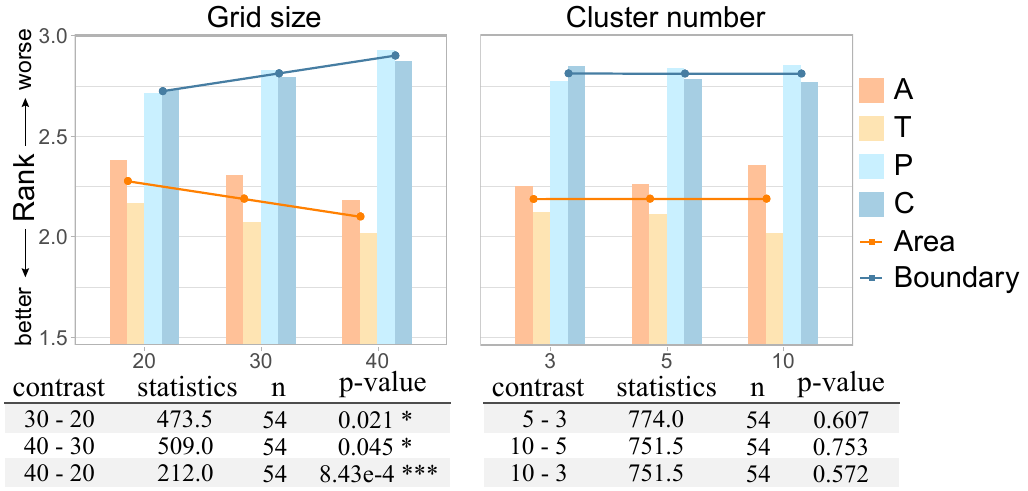}
\caption{Compare the ranks of convexity measures for different grid sizes and different cluster numbers.
{The grid size influences people’s perception of convexity, while the cluster number does not.
The tables present the results of the pairwise Wilcoxon signed-rank test on the area rank.}
Significance levels are denoted by asterisks: * indicates $p<0.05$,** indicates $p<0.01$, and *** indicates $p<0.001$.}
\label{fig:barchart}
\end{figure}

\noindent \HP2: The grid size {influences} people's perception of convexity {(confirmed)}.
{
Initially, we considered conducting a two-way ANOVA test to analyze the effects of the grid size and cluster number, given that there are two independent variables.
The interaction effect between the two variables was not significant in the ANOVA test, suggesting that their effects can be analyzed separately.
Additionally, our data violated the normality assumption.
To analyze these two variables separately, we utilized the nonparametric Friedman test and pairwise Wilcoxon signed-rank tests, which do not depend on the normality assumption.\looseness=-1
}

We first analyzed {the effect of {the} grid size on the ranks of two representative measures, measure \textbf{T} and measure \textbf{P}, which respectively represent area-based and boundary-based measures}.
However, we did not observe any significant effect.
{After conducting a more thorough analysis, we noticed that participants who strongly preferred measure \textbf{T} and \textbf{P} consistently ranked these measures highest, and that varying the grid size had little effect on their {ranking results.}}
{Therefore, we shifted our focus to analyzing {the} area rank {(\textbf{A}+\textbf{T})/2} rather than analyzing solely the rank of measure \textbf{T}, which was a more robust approach to the analysis.}
The effect of {the} grid size on the area rank becomes significant {($\chi^2(2)=14.91,p<0.001$).
}
{In addition, we observed that the area rank consistently decreased as the grid size increased, as illustrated in~\cref{fig:barchart} left.}
{The pairwise Wilcoxon test results} further confirmed significant differences between the area rank {for} different grid sizes.
Therefore, we concluded that as the grid size increases, the area-based measure {aligns better} with people's perception.
\HP2 is confirmed.

\noindent \HP3: The cluster number {influences} people's perception of convexity {(rejected)}.
{Similarly, we conducted {the Friedman test and pairwise Wilcoxon signed-rank tests} to investigate the effect of {the} cluster number on the rank of measure \textbf{T}, the rank of measure \textbf{P}, and the area rank.}
There was no significant effect of {the} cluster number on any of these ranks. 
Additionally, {the pairwise Wilcoxon test results} showed no significant differences in area ranks between different grid sizes (\cref{fig:barchart} right).
{Based on these findings}, we concluded that the cluster number does not influence the perception of convexity.
\HP3 is rejected.

\subsection{Participant Feedback}
We analyzed participants' feedback on how they compared the convexity of different grid visualizations, and we also conducted interviews with ten participants to gather further insight into their judgment-making process.
Among the ten participants, seven preferred area-based measures, while the remaining three preferred boundary-based measures.
{Participants who {favored} area-based measures {tended to overlook} small zig-zags {along} a boundary and instead {viewed} the boundary as a line segment.}
One {faculty member} {commented} that the anti-aliasing effect in the human visual system could explain this judgment.
According to this theory, these zig-zags become less noticeable with increasing grid sizes, which is consistent with our user study findings.
Two {faculty members} with a research interest in computer graphics mentioned that they tended to compare the grid visualizations with the corresponding Voronoi diagram{s} when making judgments.
They regarded a significant difference between the two as a sign of poor convexity.
{This explains why they do not like the layout optimized for boundary-based measures: a shape \glyph{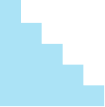} tends to become \glyph{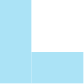} during the optimization of boundary-based measures, resulting in a larger difference compared to the corresponding Voronoi diagram.}
In contrast, the participants who preferred boundary-based measures usually paid more attention to the number of zig-zags on the boundary.
They generally preferred a cluster shape with fewer edges along its boundary.

\section{Cluster-Aware Grid Layout Method}
\label{sec:method}

\begin{figure}[!htb]
\centering
\includegraphics[width=\linewidth]{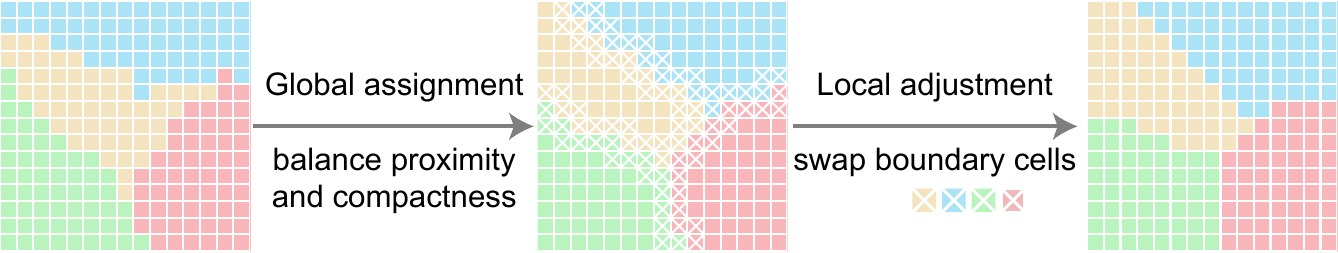}
\caption{{Our layout pipeline: our method first balances proximity and compactness in the global assignment phase, and then improves the convexity by swapping the boundary cells in the local adjustment phase}.}
\vspace{-3mm}
\label{fig:pipeline}
\end{figure}

\subsection{Design Principles}
\label{sec:principle}
Our cluster-aware grid layout method aims to {improve the analysis efficiency by enhancing the recognition of clusters~\cite{chen2021interactive,kehlbeck2021sp}}.
This is well aligned with the Gestalt principles of perceptual grouping~\cite{todorovic2008gestalt,rottmann2022mosaicsets,wagemans2012century,lu2019winglets}.
These principles {investigate} how certain elements tend to be perceived as a group.
They can be classified into two categories: layout-irrelevant {principles} and layout-relevant principles. 
Layout-irrelevant principles describe how visual elements are grouped regardless of their spatial arrangement, which includes similarity, common fate, connectedness, and common region. 
In contrast, layout-relevant principles depend on the spatial relationships between elements, which include proximity, compactness, symmetry, and convexity. 
Our layout method is based on layout-relevant principles.
Since the convexity principle overrules the symmetry principle~\cite{kanizsa1976convexity},
we employ the following three principles to develop a cluster-aware grid layout method:

\noindent \textbf{Proximity}: Similar samples should be placed close to each other.

\noindent \textbf{Compactness}: Samples in the same cluster should be placed in a compact form.

\noindent \textbf{Convexity}: Samples in the same cluster should form a convex shape.\looseness=-1

\subsection{Measuring Cluster Preservation}

The key in developing our layout method is to quantify the three measures that correspond to the three selected Gestalt principles.

\noindent\textbf{Proximity}.
To preserve proximity in a grid layout, the samples with higher similarity should have smaller Euclidean distances between their corresponding cell positions.
Let $S=\{s_1,s_2,\ldots,s_n\}$ denote the samples and $\{g(s_1),g(s_2),\ldots,g(s_n)\}$ denote their corresponding cell positions.
Given two samples $s_i$ and $s_j$, their similarity is denoted as $c_{ij}\in [0,1]$, 
and their Euclidean distance {on} the grid is computed by $\lVert g(s_i)-g(s_j) \rVert$, where $\lVert\cdot\rVert$ is the Euclidean norm.
The proximity of a grid layout is then determined by:
\begin{equation}
\mathrm{Prox} = \sum\nolimits_{i=1}^{n}\sum\nolimits_{j=1}^{n}\left(w\lVert {g(s_i)-g(s_j)\rVert-(1- c_{ij})}\right)^2,
\label{eq:prox}
\end{equation}
where $w$ is a scaling factor to ensure that the first term ranges from 0 to 1.
A smaller proximity value indicates better preservation of proximity.

\noindent\textbf{Compactness}.
{Proximity does not consider the cluster information and cannot guarantee the samples within the same cluster form a compact shape.
To preserve compactness, samples should be placed closer to their corresponding cluster centers.
}
Following Rottmann's method~\cite{rottmann2022mosaicsets}, 
the compactness of the grid layout is {measured} by the total distance between the cell position of each sample and its corresponding cluster center:
\begin{equation}
\text{Comp} = \sum\nolimits_{i=1}^{n}\lVert {g(s_i) - \mu_{i}}\rVert^{2},
\label{eq:comp}
\end{equation}
where $\mu_i$ is the corresponding cluster center of sample $s_i$.
It is computed as the average cell position over the samples in the same cluster as $s_i$.
A smaller {compactness} value indicates a more compact layout.

\noindent\textbf{Convexity}.
{As described in~\cref{subsec:userstudy}, there are four representative convexity measures, \emph{area ratio}, \emph{triple ratio}, \emph{perimeter ratio}, and \emph{cut ratio}. 
The quantifying method of each measure for a polygon is also introduced in this section.
Once a user selects one of the aforementioned measures, the convexity of a grid layout is calculated as the average convexity score over all the cluster shapes in the grid layout}.

\subsection{Layout Algorithm}

\noindent\textbf{Optimization strategy}.
{Given the large search space spanning over proximity, compactness, and convexity, it is impractical to achieve a perfect balance among the three measures}.
{Our analysis of the calculation of these three measures reveals that proximity and compactness are affected by all grid cells, whereas convexity is sensitive to the {boundary cells} between different clusters}.
Based on this finding, we employ a global-local strategy {to simplify the optimization process}.
As shown in~\cref{fig:pipeline}, the global assignment achieves a good balance between proximity and compactness.
The local adjustment swaps the boundary cells between clusters to improve convexity
{without apparently affecting the achieved proximity and compactness.}
The effectiveness of the {global-local} strategy is {demonstrated} in the ablation study (\cref{table:ablation}).

\noindent\textbf{Global assignment}.
The goal of this phase is to generate {a global layout} that simultaneously optimizes proximity and compactness.
Instead of directly optimizing proximity defined in~\cref{eq:prox}, we take a grid layout generated by a proximity-preserving method as input and then minimize the distance between the input layout and our layout.
This offers the flexibility to integrate any existing proximity-preserving method into our layout {pipeline}.
Let $V=\{v_1,v_2,\ldots,v_n\}$ denote the grid cells.
Without loss of generality, we set $v_i$ as the input cell position of $s_i$.
Our layout can be viewed as a new assignment from the samples $S$ to the grid cells $V$,
{which is} denoted by a binary matrix $\bm{\delta}=\{\delta_{ij}\}_{1\le i, j\le n}$.
Here, $\delta_{ij}=1$ indicates that $s_i$ is assigned to $v_j$ (\ie, $g(s_i)=v_j$), and otherwise, $\delta_{ij}=0$.
{The proximity of the layout is measured by \shixia{its distance to the input layout}}:
\begin{equation}
\mathrm{Prox}(\bm{\delta}) = \sum\nolimits_{i=1}^{n}\lVert g(s_i)-v_i\rVert^2=\sum\nolimits_{i=1}^{n}\sum\nolimits_{j=1}^{n}\lVert {v_j-v_i \rVert^{2}}\delta_{ij}.
\label{eq:prox2}
\end{equation}
{With the notation $\bm{\delta}$, the compactness can be rewritten as:}
\begin{equation}
\mathrm{Comp}(\bm{\delta}) = \sum\nolimits_{i=1}^{n}\lVert {g(s_i)-\mu_i \rVert^{2}}=\sum\nolimits_{i=1}^{n}\sum\nolimits_{j=1}^{n}\lVert {v_j-\mu_i \rVert^{2}}\delta_{ij}.
\label{eq:comp2}
\end{equation}
{To simultaneously optimize proximity and compactness, the measures defined in~\cref{eq:prox2,eq:comp2} are combined}, and the global layout is {achieved} by minimizing
\begin{align}
\begin{split}
\underset{\bm{\delta}}{\mathrm{minimize}}\quad &\sum\nolimits_{i=1}^{n}\sum\nolimits_{j=1}^{n}\left(\lambda\lVert {v_j-v_i \rVert^{2}}+(1-\lambda)\lVert {v_j - \mu_{i}}\rVert^{2}\right)\delta_{ij},\\
\mathrm{subject\ to}\quad&\sum\nolimits_{i=1}^{n}\delta_{ij}=1,\ \forall j\in\{1,2,\ldots,n\},\\
&\sum\nolimits_{j=1}^{n}\delta_{ij}=1,\ \forall i\in\{1,2,\ldots,n\},\\
&\delta_{ij}\in\{0,1\},\ \forall i,j,
\label{eq:global}
\end{split}
\end{align}
where 
$0\le \lambda\le 1$ is the weight to {balance proximity and compactness}.
The constraints ensure a one-to-one assignment from {samples} to {grid cells}.
The optimization problem defined in~\cref{eq:global} is a linear assignment problem and can be efficiently solved with an accelerated Jonker-Volgenant algorithm~\cite{chen2020oodanalyzer}.

{Determining weight $\small \lambda$ that balances proximity and compactness is crucial for generating a global layout that achieves good results. 
Using a fixed value, such as $\small \lambda=0.5$, may not be optimal for all cases, and manually tuning the parameter is labor-intensive and requires expertise}.
{Thus,} we employ the multi-task learning method proposed by Liu~\etal\cite{Liu_Liang_Gitter_2019} to determine $\small \lambda$.
The key idea is to increase the weight of the task with poor performance {so that it can be further improved}.
To assess the performance of each task, we compare the current {proximity/compactness score with its optimal one}.
Specifically, we first obtain the layout with the optimal proximity $\small \bm{\delta}_{p}$, which is the input layout, and the layout with the optimal compactness $\small \bm{\delta}_c$, which is computed by optimizing compactness solely.
The performance of optimizing proximity is determined by $\small \Delta_{\text{Prox}} = (\text{Prox}(\bm{\delta}) - \text{Prox}(\bm{\delta}_p)) / (\text{Prox}(\bm{\delta}_c) - \text{Prox}(\bm{\delta}_p))$, and the performance of optimizing compactness is determined by $\small \Delta_{\text{Comp}} = (\text{Comp}(\bm{\delta}) - \text{Comp}(\bm{\delta}_c))/(\text{Comp}(\bm{\delta}_p) - \text{Comp}(\bm{\delta}_c))$.
Once the performance of both tasks is determined, we calculate $\small \lambda$ as $\small \Delta_{\text{Prox}} / (\Delta_{\text{Prox}}+\Delta_{\text{Comp}})$ and update the layout with the new weight.
The above procedure is repeated until the layout converges.

\noindent\textbf{Local adjustment}. 
{After obtaining the global layout, the local adjustment improves convexity by swapping the boundary cells between different clusters}.
In our implementation, a cell {is considered to be a} boundary cell if at least one of its neighboring cells within a 3x3 region {belongs to} a different cluster.
{We only swap boundary cells since they directly impact convexity. 
At each iteration, a boundary cell is randomly selected, and all possible swaps between the selected cell and other boundary cells are enumerated.
The convexity is evaluated after each swap, and the optimal swap that increases convexity most is chosen.}
The iterative process stops when all the boundary cells are processed.\looseness=-1

\noindent\textbf{Supporting hierarchical grid layout}.
{The two-phase layout method creates a flat grid layout that simultaneously optimizes} proximity, compactness, and convexity.
When handling a large number of samples, a hierarchical grid layout is necessary to support level-of-detail exploration.
If the input is a hierarchical layout, 
our method enhances compactness and convexity at each level of the hierarchy {while preserving proximity}.
Otherwise, our method creates a hierarchy using the {sampling-based technique} described in Chen~{\etal}'s work~\cite{chen2020oodanalyzer}.
It first samples a set of representative samples and {creates} the grid layout at the top level.
{The remaining samples are assigned to their closest representative sample}.
When the user selects a sub-region for exploration, the selected samples and some of the samples assigned to them are used to generate a grid layout in the same way.
For both methods, we try to preserve the relative positions of the previously displayed samples for maintaining the mental map during exploration.\looseness=-1

\section{Evaluation}
\label{sec:evaluation}

\subsection{Quantitative Evaluation}

\subsubsection{Datasets and Experimental Settings}
\label{sec:settings}

\begin{table*}[!t]
\fontsize{8}{8}\selectfont
\setlength{\tabcolsep}{1.25em}
\centering
\caption{Comparison of six measures of all the methods. Baseline: OoDAnalyzer~\cite{chen2020oodanalyzer}; G: global; L: local; T: triple ratio; P: perimeter ratio.}
\vspace{-1.5mm}
\begin{tabular}{lrrrrrrrr}
\toprule
Measure    & Baseline & Ours-G & Ours-L(T) & Ours-L(P) & Ours-L(T)-G & Ours-L(P)-G & Ours-G-L(T) & Ours-G-L(P)  \\
\midrule
Proximity        & \textbf{1.000} & 0.996 & 0.998 & 0.992 & 0.997 & 0.996 & 0.996 & 0.994 \\
Compactness        & 0.964 & \textbf{0.970} & 0.967 & 0.963 & 0.969 & \textbf{0.970} & \textbf{0.970} & 0.968 \\
Area ratio        & 0.669 & 0.882 & 0.900 & 0.852 & 0.869 & 0.883 & \textbf{0.913} & 0.895 \\
Triple ratio         & 0.936 & 0.991 & 0.995 & 0.954 & 0.989 & 0.991 & \textbf{0.997} & 0.978 \\
Perimeter ratio        & 0.812 & 0.834 & 0.857 & 0.926 & 0.833 & 0.835 & 0.866 & \textbf{0.935} \\
Cut ratio      & 0.802 & 0.870 & 0.872 & 0.913 & 0.866 & 0.872 & 0.890 & \textbf{0.934} \\
\bottomrule
\end{tabular}
\vspace{-2mm}
\label{table:ablation}
\end{table*}

\begin{table*}[!t]
\fontsize{8}{8}\selectfont
\setlength{\tabcolsep}{.3em}
\centering
\caption{Comparison of six measures of the baseline (OoDAnalyzer~\cite{chen2020oodanalyzer}), Ours-T, and Ours-P with 3 different grid sizes.}
\vspace{-2mm}%
\begin{tabular}{c|ccc|ccc|ccc|ccc|ccc|ccc}
\toprule
\multirow{2}{*}{Grid size} & \multicolumn{3}{c|}{Proximity}  & \multicolumn{3}{c|}{Compactness} & \multicolumn{3}{c|}{Area ratio}  & \multicolumn{3}{c|}{Triple ratio} & \multicolumn{3}{c|}{Perimeter ratio} & \multicolumn{3}{c}{Cut ratio} \\
& Basel. & Ours-T & Ours-P & Basel. & Ours-T & Ours-P & Basel. & Ours-T & Ours-P & Basel. & Ours-T & Ours-P & Basel. & Ours-T & Ours-P & Basel. & Ours-T & Ours-P \\
\midrule
20x20 & \textbf{1.000} & 0.996 & 0.993 & 0.965 & \textbf{0.970} & 0.968 & 0.750 & \textbf{0.898} & 0.896 & 0.951 & \textbf{0.995} & 0.976 & 0.843 & 0.875 & \textbf{0.938} & 0.839 & 0.895 & \textbf{0.935} \\
30x30 & \textbf{1.000} & 0.996 & 0.994 & 0.964 & \textbf{0.970} & 0.969 & 0.664 & \textbf{0.916} & 0.897 & 0.938 & \textbf{0.997} & 0.979 & 0.816 & 0.865 & \textbf{0.935} & 0.801 & 0.891 & \textbf{0.935} \\
40x40 & \textbf{1.000} & 0.995 & 0.993 & 0.962 & \textbf{0.970} & 0.969 & 0.591 & \textbf{0.926} & 0.893 & 0.919 & \textbf{0.998} & 0.978 & 0.775 & 0.858 & \textbf{0.933} & 0.766 & 0.885 & \textbf{0.932} \\
\bottomrule
\end{tabular}
\vspace{-4mm}
\label{table:full}
\end{table*}

\noindent\textbf{Datasets}.
We evaluated the effectiveness of our layout {method} on 11 datasets.
Ten of them (Animals, Cifar10, Indian Food, Isolet, MNIST, Stanford Dogs, Texture, USPS, Weather, Wifi) are from Xia~\etal's work~\cite{xia2022interactive}, while an additional dataset, OoD-Animals, is from {OoDAnalyzer for data quality analysis}~\cite{chen2020oodanalyzer}.
There are eight image datasets, two textual datasets (Isolet and Texture), and one tabular dataset (Wifi).
{More details of these datasets are in supplemental material.}
For images, we used CLIP~\cite{radford2021learning}, a state-of-the-art pre-trained model to extract the feature vectors.
For textual data and tabular data, we used the feature vectors provided by the dataset.
{The cluster labels of each dataset were set as the predictions, which were obtained using the $k$-NN classifier, where $k$ was determined using cross-validation.}

\noindent\textbf{Experimental settings}.
{The baseline is the state-of-the-art proximity-preserving grid layout method proposed by Chen~\etal~\cite{chen2020oodanalyzer}.}
{The method first projects samples into a 2D space using t-SNE, and then assigns the projected samples to cells by solving a linear assignment problem}.
Ours-G improves compactness only through the \textbf{g}lobal assignment, while Ours-L(T) and Ours-L(P) only use \textbf{l}ocal adjustment to improve \textbf{t}riple ratio (area-based) or \textbf{p}erimeter ratio (boundary-based).
{We chose these two measures because they were representative of area-based measures and boundary-based measures, respectively.}
We also evaluated the performance of different combination orders of the global phase and local phase, which resulted in four more methods: Ours-L(T)-G, Ours-L(P)-G, Ours-G-L(T), and Ours-G-L(P).
In the following experiments, we generate the grid layouts using 3 different grid sizes: 20x20, 30x30, and 40x40, consistent with the grid sizes used in our user study.

\noindent\textbf{Evaluation criteria}.
We used {six measures to evaluate the quality of the grid layout: proximity, compactness, {and four convexity measures, including} triple ratio, area ratio, perimeter ratio, and cut ratio.} 
{The scores of the four convexity measures} range from 0 to 1, and a higher score indicates better convexity,
whereas the proximity and compactness scores ({\cref{eq:prox2,eq:comp2}}) range from 0 to infinity, and a smaller score indicates better proximity/compactness.
To facilitate comparison, 
{we apply the transformation $x\mapsto \exp(-x)$ to the proximity and compactness scores, such that they are normalized to a range of 0 to 1,} and a higher score indicates better proximity/compactness.

\subsubsection{Ablation Results}
\label{subsec:ablation}
{Our study was designed to examine the impacts of both the global assignment phase and the local adjustment phase.
The effectiveness was demonstrated by} comparing the associated methods using the six measures. 
The results presented in \cref{table:ablation} were averaged over the 11 datasets and 3 grid sizes, and full results are available in supplemental material.
All our methods {preserved} proximity well, with scores above 0.99. 
Thus, our analysis mainly focused on compactness and convexity. 
Compared with the baseline, Ours-G improved compactness from 0.964 to 0.970.
{Moreover, it also {showed} improvement on all four convexity measures, with more notable improvement on the area-based ones (area ratio and triple ratio)}.
This is because the optimization of compactness leads to regular cluster shapes, which have higher scores in area-based measures.
Regarding convexity measures, Ours-L(T) performed better than baseline/Ours-G/Ours-L(P) in terms of the area ratio and triple ratio, while Ours-L(P) surpassed baseline/Ours-G/Ours-L(T) with respect to the perimeter ratio and cut ratio.
This indicates that optimizing an area-based convexity measure can lead to a large improvement on the other area-based measures due to their related optimization goals. 
The same applies to boundary-based measures.

{Moreover, our ablation study explored the optimal order to combine the global assignment phase and the local adjustment phase.}
It was observed that the results of Ours-G, Ours-L(T)-G, and Ours-L(P)-G were {quite} similar.
This indicates that the changes made in the global assignment phase have a greater influence than those in the local adjustment phase. 
Therefore, the local adjustment should be conducted after the global assignment.
The comparison between Ours-G-L(T)/Ours-G-L(P) and Ours-G reveals that the local adjustment phase further improves convexity without apparently affecting the proximity and compactness achieved in the global assignment phase.
Moreover, Ours-G-L(T) performed better than Ours-L(T) in terms of the area ratio and triple ratio, and 
Ours-G-L(P) {performed better than Ours-L(P) regarding the perimeter ratio and cut ratio.}
This indicates that the global assignment phase provides a better initial layout for the local adjustment phase, leading to a larger improvement on convexity measures.
Therefore, we choose Ours-G-L(T) and Ours-G-L(P) as the primary methods, which are abbreviated as \textbf{Ours-T} and \textbf{Ours-P} in the following experiments.

\begin{figure}[!htb]
\centering
\includegraphics[width=\linewidth]{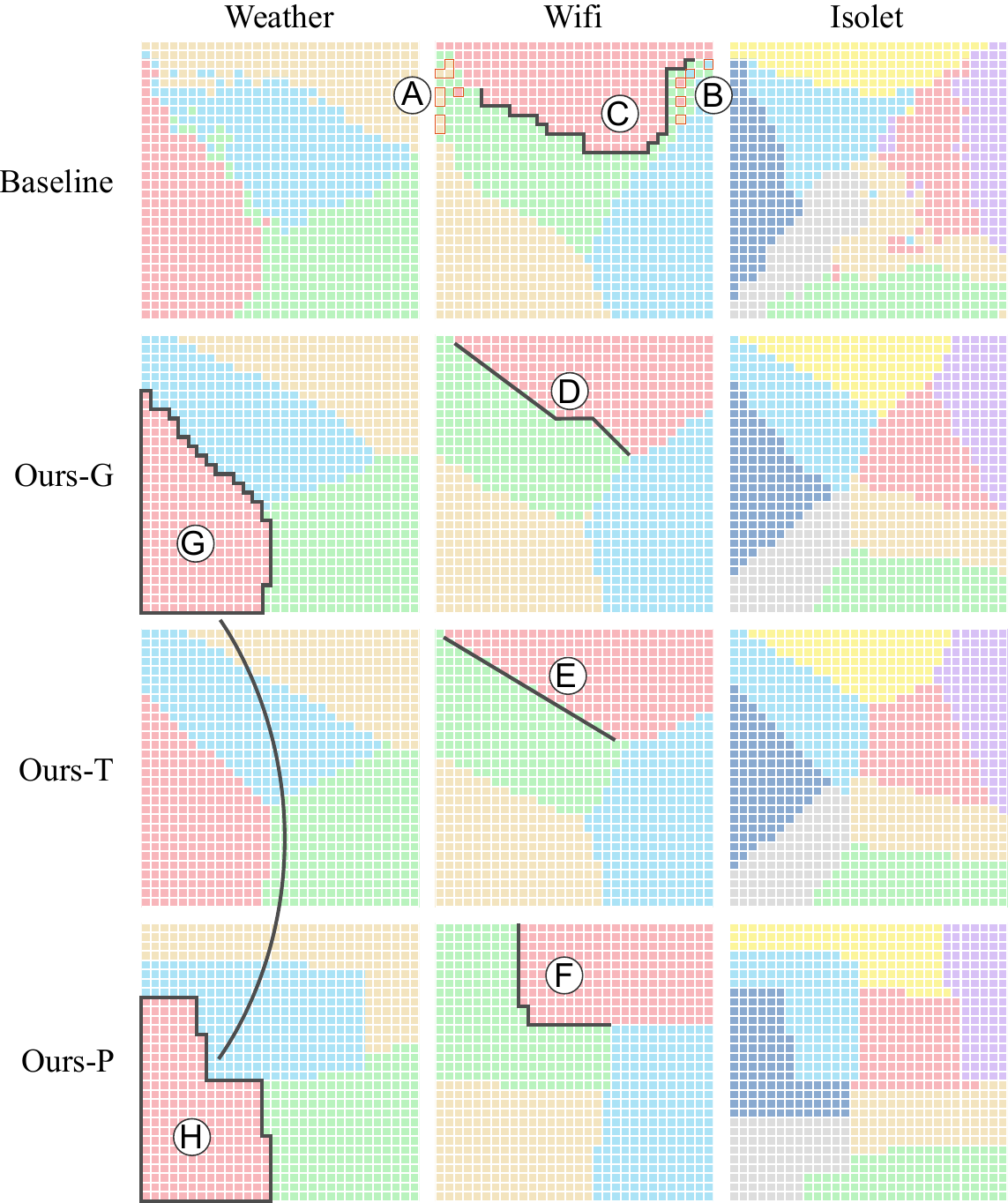}
\caption{Comparison of the layouts generated by the baseline (OoDAnalyzer~\cite{chen2020oodanalyzer}), Ours-G, Ours-T, and Ours-P.
{Using our methods, samples that fall into other clusters (A and B) are placed into their corresponding clusters, and irregular boundaries between different clusters (C) become regular (D, E, and F). Boundaries in Ours-G and Ours-T contain many slanted segments (D, E, and G), while boundaries in Ours-P mainly consist of horizontal and vertical segments (F and H).\looseness=-1}}
\vspace{-4mm}
\label{fig:vis}
\end{figure}

\subsubsection{Comparison Results}
\cref{table:full} provides a detailed comparison between the baseline and Ours-T/Ours-P with 3 different grid sizes.
The reported results were averaged over 11 datasets.
In terms of proximity and compactness, there were no significant differences across different grid sizes.
Ours-T achieved the highest compactness without affecting proximity too much.
Compared to Ours-T, Ours-P achieved relatively lower scores in both proximity and compactness, but the differences were very small.
Regarding convexity measures, Ours-T performed best in terms of area-based measures, while Ours-P performed best in terms of boundary-based measures.
This indicates that our methods can fulfill the requirements of individuals who prefer either area-based or boundary-based measures.
When comparing the convexity scores across different grid sizes, it is notable that while the convexity scores achieved by the baseline method decrease as the grid size increases, the scores of area-based measures achieved by Ours-T increase with larger grid sizes.
After analyzing the calculation of area-based measures, it is discovered that the small zig-zags on the boundaries between different clusters have less impact on area-based measures as the grid size increases.
As a result, Ours-T achieved higher area ratios and triple ratios with larger grid sizes.
In contrast, the calculation of boundary-based convexity measures is not affected by the changes in grid size.
Therefore, Ours-P did not achieve higher perimeter ratios or cut ratios with larger grid sizes.

{\cref{fig:vis} presents the layout results generated by the baseline, Ours-G, Ours-T, and Ours-P on six example datasets.
The full results are summarized in supplemental material}.
In the layouts generated by the baseline, some samples fall into the clusters they do not belong to (\cref{fig:vis}A and B), and the boundaries between different clusters are irregular (\cref{fig:vis}C).
After the global assignment phase, the samples within the same cluster are grouped together, and the boundaries become more regular (\cref{fig:vis}D).
It is also noted that the results generated by Ours-G are similar to Ours-T, which is consistent with our findings that Ours-G achieves a more notable improvement on the area-based convexity measures in~\cref{subsec:ablation}.
Further comparison between Ours-G and Ours-T reveals that the boundaries generated by Ours-T are closer to line segments (\cref{fig:vis}E) than Ours-G, leading to higher area-based convexity scores.
In contrast, the boundaries generated by Ours-P mainly consist of horizontal and vertical line segments (\cref{fig:vis}F), making the layout results dissimilar from the results generated by Ours-G and Ours-T.
It is also observed that a shape \glyph{shape1} (\cref{fig:vis}G) tends to be optimized towards \glyph{shape2} (\cref{fig:vis}H), which has higher boundary-based convexity scores but lower area-based convexity scores.

\subsubsection{Running Time}

\begin{table}[!b]
\setlength{\tabcolsep}{2em}
\centering
\caption{Running time comparison (in seconds) of different methods to generate grid layouts with different sizes. {Baseline: OoDAnalyzer~\cite{chen2020oodanalyzer}}.}
\vspace{-1mm}
(a) Adaptive $\lambda$ in the global assignment phase.\\
\vspace{1mm}
\begin{tabular}{ccccc}
\toprule
Method & 20x20 & 30x30 & 40x40  \\
\midrule
Ours-G & 0.059 & 0.396 & 1.821  \\
Ours-T & 0.101 & 0.709 & 3.050  \\
Ours-P & 0.133 & 0.636 & 2.465  \\
\bottomrule
\end{tabular}
\\
\vspace{3mm}
(b) Fixed $\lambda$ in the global assignment phase.\\
\vspace{1mm}
\begin{tabular}{ccccc}
\toprule
Method &  20x20 & 30x30 & 40x40 \\
\midrule
{Baseline} & {0.013} & {0.081} & {0.319} \\ 
\midrule
Ours-G & 0.009 & 0.064 & 0.291  \\
Ours-T & 0.055 & 0.377 & 1.513  \\
Ours-P & 0.092 & 0.309 & 0.961  \\
\bottomrule
\end{tabular}
\label{table:time}
\end{table}

We evaluated the running time of our methods {for 3 grid sizes on a desktop PC with an Intel i9-13900K CPU (5.0 GHz).
The results were averaged over 11 datasets.}
As shown in~\cref{table:time}(a), our methods generated a 30x30 grid layout in less than 1 second, and a 40x40 grid layout in around 3 seconds.
Moreover, the results indicated that the global assignment phase consumed most of the time.
{Further analysis revealed that, on average, the optimal value of $\lambda$ required solving the linear assignment problem approximately six times}.
This process can be accelerated if $\lambda$ is fixed.
{A practical way to find an appropriate $\lambda$ is to test different values on a set of representative datasets and choose the one that works well for most of them.}
As shown in~\cref{table:time}(b), {the running time of Ours-G approximately equals that of the baseline} when $\lambda$ is fixed,
{and the total time of our methods (Ours-T and Ours-P) is a little bit longer. For example, when generating a 40x40 grid layout, the baseline takes 0.319 seconds, and Ours-P and Ours-T take 0.961 and 1.513 seconds, respectively.}
Previous research has shown that a minimum size of 32x32 pixels is required to identify important objects in an image~\cite{torralba2009many}.
Taking a display with a resolution of 1920x1080 as an example, each cell in a 40x40 grid has a maximum width of only $1080/40=27$ pixels, which is below the minimum size.
Thus, our methods closely approach the real-time requirements of most applications.\looseness=-1

\subsection{Use Cases}
We present two use cases to showcase how our layout method facilitates 1) identifying misclassified samples; and 2) analyzing out-of-distribution (OoD) samples.

\begin{figure}[!tb]
\centering
{\includegraphics[width=\linewidth]{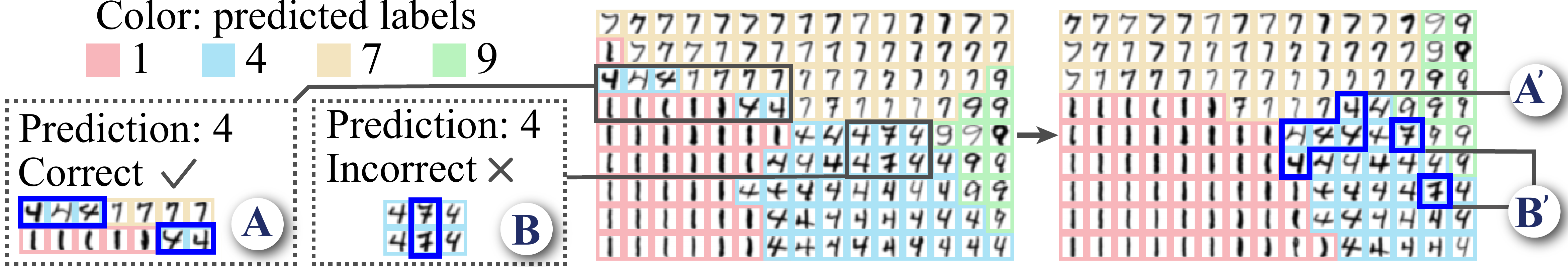}}
\put(-145,-7){(a) Baseline}
\put(-65,-7){(b) Ours-T}
\vspace{-3mm}
\caption{{Compared with the baseline (a), the samples predicted as ``4'' (A and B) are merged into the cluster of ``4'' in our layout (b).}}
\vspace{-3mm}
\label{fig:usecase1}
\end{figure}

\subsubsection{Identifying misclassified samples}
This use case {illustrates how our method aids in identifying} misclassified samples in a classification task.
We used the USPS dataset~\cite{radford2021learning}, which consists of 9,298 gray-scale images of handwritten digits from 0 to 9. 
As in the quantitative experiments, feature vectors were extracted using CLIP~\cite{radford2021learning}, and predictions were generated using a $k$-NN classifier (accuracy: 93.27\%).
Four classes, 1/4/7/9, were chosen for analysis because most of the misclassifications happened among the samples of these four classes.
{As the ground-truth labels are not available, users need to examine the samples in the grid layout to identify misclassified samples.
To facilitate the identification, samples with the same predictions were treated as a single cluster, and 
we utilized both position and color to encode cluster structures.}

{Figs.~\ref{fig:usecase1}(a) and (b) show a part of the grid layouts generated by the baseline and Ours-T, respectively.
We chose Ours-T because the triple ratio is the most favored measure in our user study.
In the baseline layout (\cref{fig:usecase1}(a)), the samples of ``4'' in \cref{fig:usecase1}$\text{A}$ are placed far away from the cluster of ``4.''
This arrangement may lead users to draw the wrong conclusion that the model predicts these samples to be not similar to ``4.''
However, these samples are similar to other samples of ``4'' and are predicted as ``4.''
By preserving cluster structures (\cref{fig:usecase1}(b)), these samples are merged into the cluster of ``4'' (\cref{fig:usecase1}$\text{A}'$), which reduces false inferences.
Furthermore, misclassified samples,
such as the samples in \cref{fig:usecase1}$\text{B}$, which are ``7'' but misclassified as ``4,''
still remain in the cluster of ``4'' and can be easily identified in the cluster-aware layout.
}

\subsubsection{Analyzing OoD samples}
The second use case illustrates how our method facilitates the analysis of OoD samples, the test samples that are not well covered by training samples.
Analyzing why OoD samples appear and adding corresponding samples to the training data can boost model performance~\cite{chen2020oodanalyzer}.
A recent work, OoDAnalyzer~\cite{chen2020oodanalyzer}, utilizes a grid visualization to help analyze OoD samples.
We were interested in whether our cluster-aware grid layout could further improve analysis efficiency and help identify more OoD samples. 
Therefore, we invited one author of OoDAnalyzer (\E1) to conduct the analysis on the OoD-Animals dataset, which was used in the case study of OoDAnalyzer.
This dataset contains 7,270 training samples and 19,413 test samples of five categories: cat, dog, rabbit, tiger, and wolf.
Following OoDAnalyzer, the color hue encodes the prediction.
The color saturation encodes the OoD score, and a darker color indicates a larger OoD score that warrants examination.
{Similar to the first use case, samples with the same predictions were treated as a single cluster.}

\E1 first compared the grid layout used in OoDAnalyzer (\cref{fig:teaser}(c)) and the corresponding grid layout generated by Ours-T (\cref{fig:teaser}(d)).
{The OoD samples found in the previous work~\cite{chen2020oodanalyzer} can be easily identified in our layout.
For example,}
in \cref{fig:teaser}(c), some dark blue cells (cat) fell into different regions $\text{C}_1\text{--}\text{C}_3$.
After examining the samples in these three regions one by one, \E1 found that all these dark blue cells {were} the samples of ``leopard.''
Since there was no category ``leopard'' in the training data, these samples were predicted as {the most similar category} ``cat'' (blue) but had high OoD scores.
However, it was hard to analyze them in OoDAnalyzer because they were scattered into several categories.
{This arrangement may even lead users to draw the wrong conclusion that the model predicts the samples in $\text{C}_3$ to be closely related to ``rabbit.''}
By preserving cluster structures (\cref{fig:teaser}(d)), {these OoD samples of ``leopard''} were grouped together in {the cluster of ``cat''} ($\text{C}'$).
{This enables users to efficiently identify these OoD samples as a whole and take the associated actions.}\looseness=-1

\begin{figure}[!tb]
\centering
{\includegraphics[width=\linewidth]{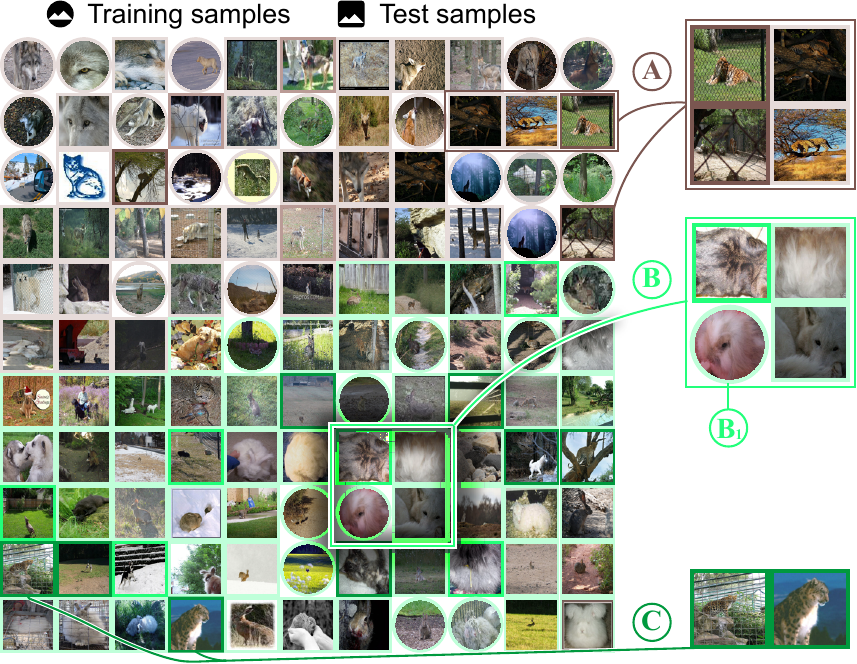}}
\vspace{-3mm}
\caption{{After zooming into region~\cref{fig:teaser}D, more samples with high OoD scores are identified (A and C). Our layout also reveals that some samples with only part of an animal (B) are misclassified as ``rabbit'' due to a low-quality training sample with only the hair of a rabbit (B${}_1$).}}
\vspace{-3mm}
\label{fig:usecase}
\end{figure}

{Our cluster-aware layout also helped identify more OoD samples that were not identified in the previous work~\cite{chen2020oodanalyzer}.
For example,} the samples with high OoD scores in \cref{fig:teaser}D were not identified before because they were not grouped together in OoDAnalyzer (\cref{fig:teaser}(c)) and thus did not trigger examinations.
\E1 then zoomed into \cref{fig:teaser}D to examine these samples (\cref{fig:usecase}).
Upon examination, \E1 discovered that the samples with high OoD scores were ``leopard'' or ``tiger'' but predicted as ``wolf'' (\cref{fig:usecase}A) or ``rabbit'' (\cref{fig:usecase}C).
Furthermore, during the examination, a new finding was discovered by \E1.
Many samples that only included part of an animal were predicted as ``rabbit'' (\cref{fig:usecase}B).
{Upon further analysis, it was discovered that a low-quality training sample with only the hair of a rabbit (\cref{fig:usecase}$\text{B}_1$) explained why the model tended to recognize images with hair as ``rabbit.''}

\subsection{Expert Feedback and Discussions}
{We interviewed three experts who are not co-authors of this work}.
\E1 participated in the second use case,
\E2 is a senior computer vision researcher who often utilizes the grid layout to explore image datasets, and \E3 is a mathematician with rich experience in convexity.
Initially, we {introduced} both OoDAnalyzer and its enhanced version {that integrates Ours-T as the layout method}. 
Then the experts freely explored the OoD-Animals dataset using both of the systems and compared the layouts.
We also collected their feedback during the exploration process.
Each interview lasted 40-60 minutes.
{Our layout method received positive feedback from all the experts regarding its usefulness}.

\noindent\textbf{Enhancing cluster perception}.
All the experts agreed that our layout method enhanced the cluster perception and aided in efficiently identifying confused sample predictions.
\E2 commented that some light colors used in the original {OoDAnalyzer} system were hard to differentiate from each other, such as light brown \lightbrownbox\thinspace and light purple \lightpurplebox.
As a result, it took him more time to recognize the predictions when the samples with similar prediction colors were mixed together.
Our method alleviated this by grouping samples with the same predictions together.
\E1 indicated that enhancing cluster perception helped him find more OoD samples.
{``As the samples with high OoD scores fall into different regions in the original system,
I missed some of them in my previous analysis.
The clear cluster structures in the new layout enable me to identify more OoD samples easily and prepare better training data.''}\looseness=-1

\noindent\textbf{Extensibility}.
The extensibility of our layout method comes from three sources.
First, our method takes a grid layout as input, which offers the flexibility to integrate any existing layout method into our layout process.
In addition to an input grid layout, \E1 pointed out that our method could generate a grid layout based on a scatterplot {by modifying the proximity calculation}.
Therefore, existing dimensional reduction techniques are readily integrated into our method for visualizing high-dimensional data.
Second, our method supports different convexity measures to meet different analysis {needs.
For example, optimizing area-based measures usually makes fewer adjustments and hence better preserves both the achieved proximity and compactness. Optimizing boundary-based measures tends to result in cluster shapes that are close to the combinations of rectangles, making them suitable for smaller grid sizes.}
Users can either choose a convexity measure from our provided measures or even customize a measure that fits their goals.
Third, the global-local strategy can be extended to optimize other measures, such as aesthetics and continuity.
The global assignment can be used to optimize measures that are affected by all the grid cells, while the local adjustment can iteratively optimize measures that are only affected by specific cells.
{For example, E3 noted that when continuity is a concern, it is necessary to reject a swap operation that separates a cluster into two disconnected parts.}

{The experts have also suggested two interesting research topics, which provide insights for future studies.}

\noindent\textbf{Combining with other design variables}.
We have shown the effectiveness {of our method} in enhancing the perception of cluster structures by adjusting the positions of visual elements.
{\E1 pointed out that in addition to the positions, there were also other} design variables that could be used to enhance the cluster perception, such as color and shape.
Following the Gestalt principle of similarity~\cite{wagemans2012century}, we can encode samples in the same cluster using the same color or shape.
{This is already employed in our method.}
The principle of common region also suggests that we can add contours to group the samples in the same cluster.
To further improve cluster perception, it is interesting to investigate how to combine these methods effectively. 
In addition, combining multiple methods together may cause cognitive overload.
Therefore, it is worth studying how to balance the trade-off between enhancing cluster perception and avoiding cognitive overload.

\noindent\textbf{Interactive editing}.
Although our method has provided a grid layout that well preserves cluster structures, {\E2} expressed his need to further adjust the grid layout based on his requirements.
``Sometimes I would like to locally modify the boundary to make it clearer or force two similar samples to be placed adjacently.''
It is worth exploring user-friendly interactions that allow users to directly edit the layouts toward their desired effects.
{For example,} at the sample level, we consider supporting users to move samples to the desired positions and select multiple samples to add must-link/cannot-link constraints among them.
At the cluster level, users can {sketch} the desired cluster shapes or change the convexity calculation of certain clusters.
At the global level, {users {have the flexibility to}} adjust $\lambda$, {which balances the preservation of the original layout and the preservation of cluster structures.
The larger the $\lambda$ value, the better the original layout is preserved.
}
\looseness=-1

\section{Conclusion}
\label{sec:conclusion}
We present a cluster-aware grid layout method that enhances the perception of cluster structures by optimizing proximity, compactness, and convexity simultaneously.
Starting from the input layout generated by any existing method, our method first optimizes proximity and compactness together in the global assignment phase.
Then, a local adjustment phase swaps boundary cells between different clusters to improve convexity.
To determine the convexity measure used in the local adjustment phase, we conducted a user study and {identified} two {representative} measures, triple ratio and perimeter ratio, to accommodate the diverse preferences of users.
The quantitative evaluations demonstrate that our method achieves {experimentally} optimal balances among proximity, compactness, and convexity.
Two use cases are also presented to demonstrate {how our method can be practically useful in exploring image datasets and analyzing model predictions}.

	
\acknowledgments{%
This work was supported by the National Natural Science Foundation of China under grants U21A20469, 61936002, the National Key R\&D Program of China under Grant 2020YFB2104100, grants from the Institute Guo Qiang, THUIBCS, and BLBCI, and in part by Tsinghua-Kuaishou Institute of Future Media Data.
The authors would like to thank Yifan Hu, Zhen Li, Zhaowei Wang, and Jun Yuan for their valuable contributions to the discussions, and Jiangning Zhu for his assistance in proofreading and voicing our video.
}

\bibliographystyle{abbrv-doi-hyperref}

\bibliography{reference}

\end{document}


\maketitle

\section{User study details}
\label{sec:user-study-supple}
A user study is presented in Section 3.1 of the paper.
We report here the content of the tutorial shown to participants, the trials used in practice session and formal study and the questionnaire.

\subsection{Tutorial and practice trials }
\label{subsec:tutorial}
We present a tutorial video to participants that introduced the definition of convex polygons and the user interface of the study system. Here is the content of the tutorial.
\hspace*{\fill}\\

\begin{figure}[!h]
\centering
\includegraphics[width=0.65\linewidth]{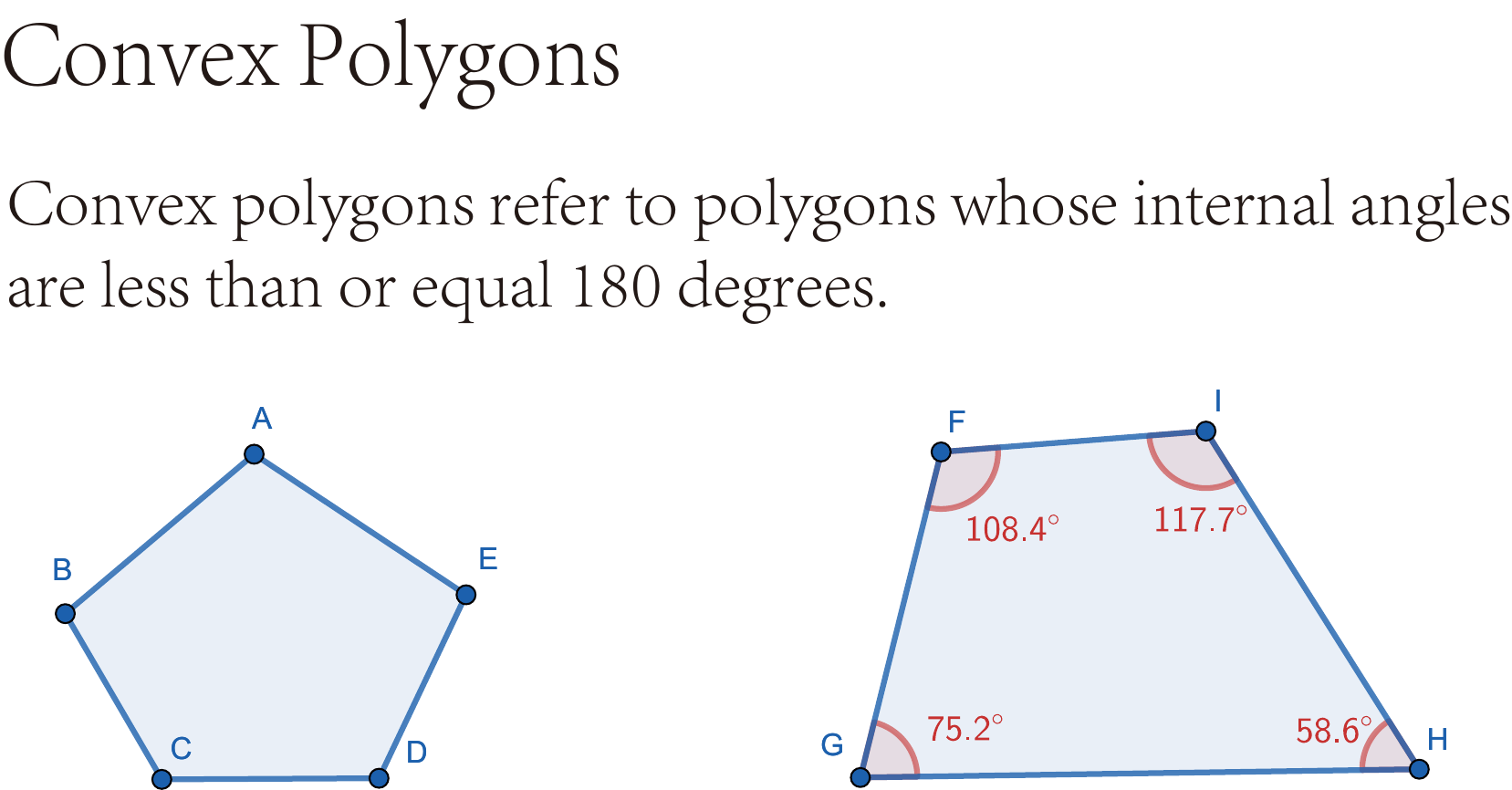}
\hspace*{5mm}
\caption{Convex polygons.}
\label{fig:tutorial1}
\end{figure}

\begin{figure}[!h]
\centering
\includegraphics[width=0.72\linewidth]{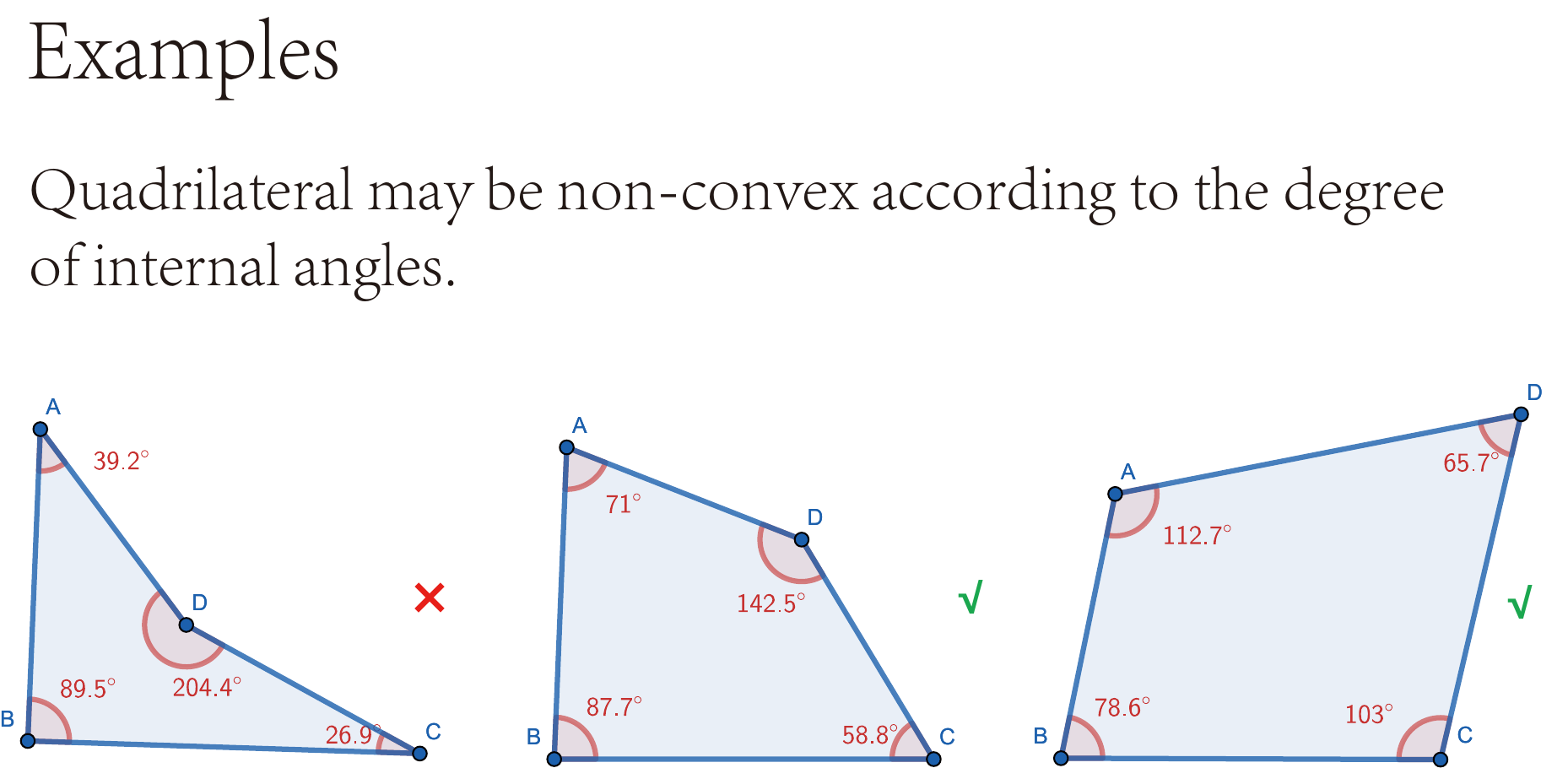}
\caption{Example of polygons.}
\label{fig:tutorial2}
\end{figure}

\noindent\textbf{What is a convex polygon?}
Convex polygons refer to polygons whose internal angles are less than or equal 180 degrees.(\cref{fig:tutorial1})
Triangles are convex polygons because all internal angles are less than 180 degrees.
Besides, regular polygons are always convex.
However, quadrilateral may be non-convex according to the degree of internal angles.(\cref{fig:tutorial2})

\begin{figure}[!h]
\centering
\includegraphics[width=0.72\linewidth]{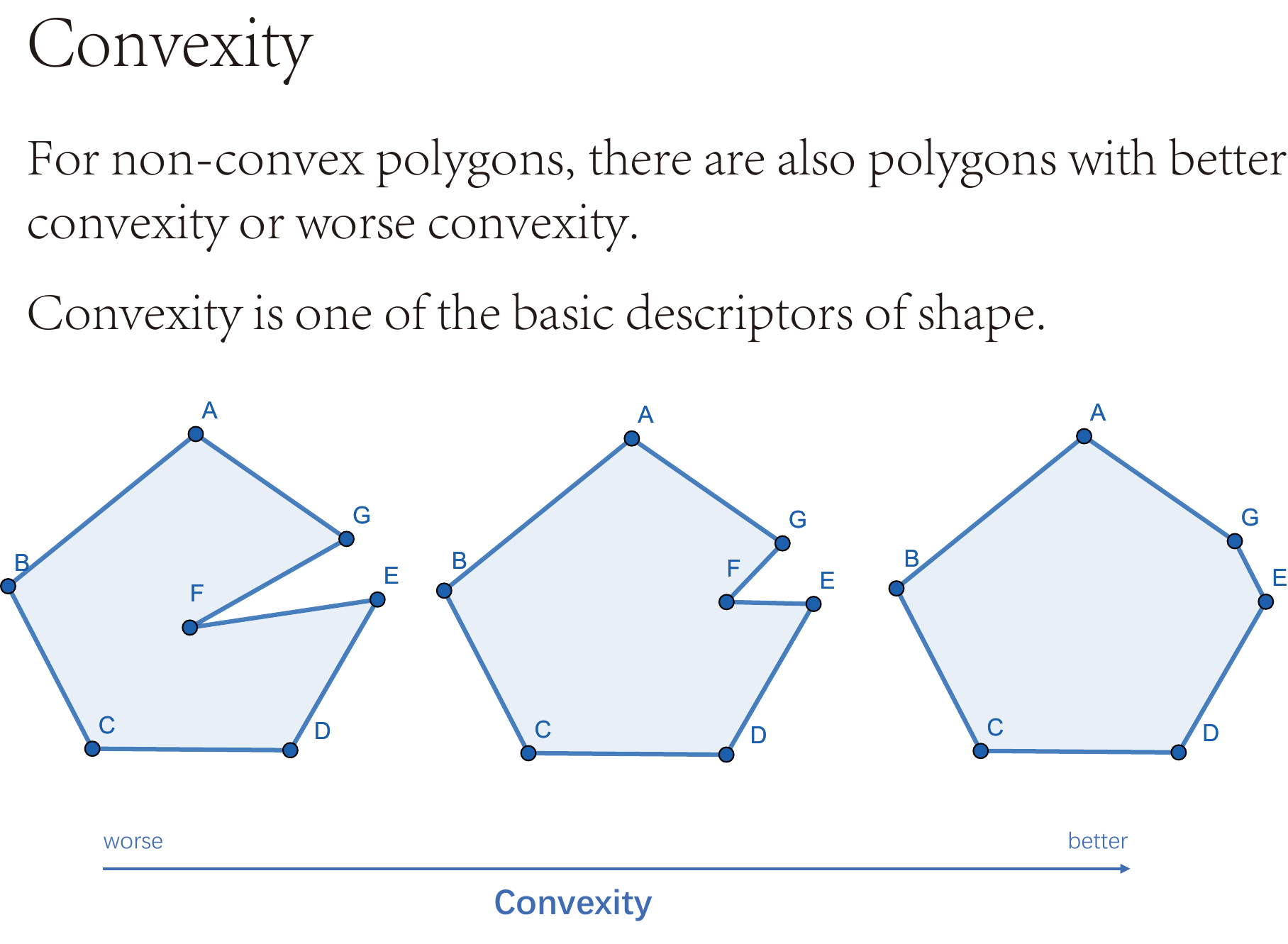}
\hspace*{1mm}
\caption{Convexity.}
\label{fig:tutorial3}
\end{figure}

\begin{figure}[!h]
\centering
\includegraphics[width=0.72\linewidth]{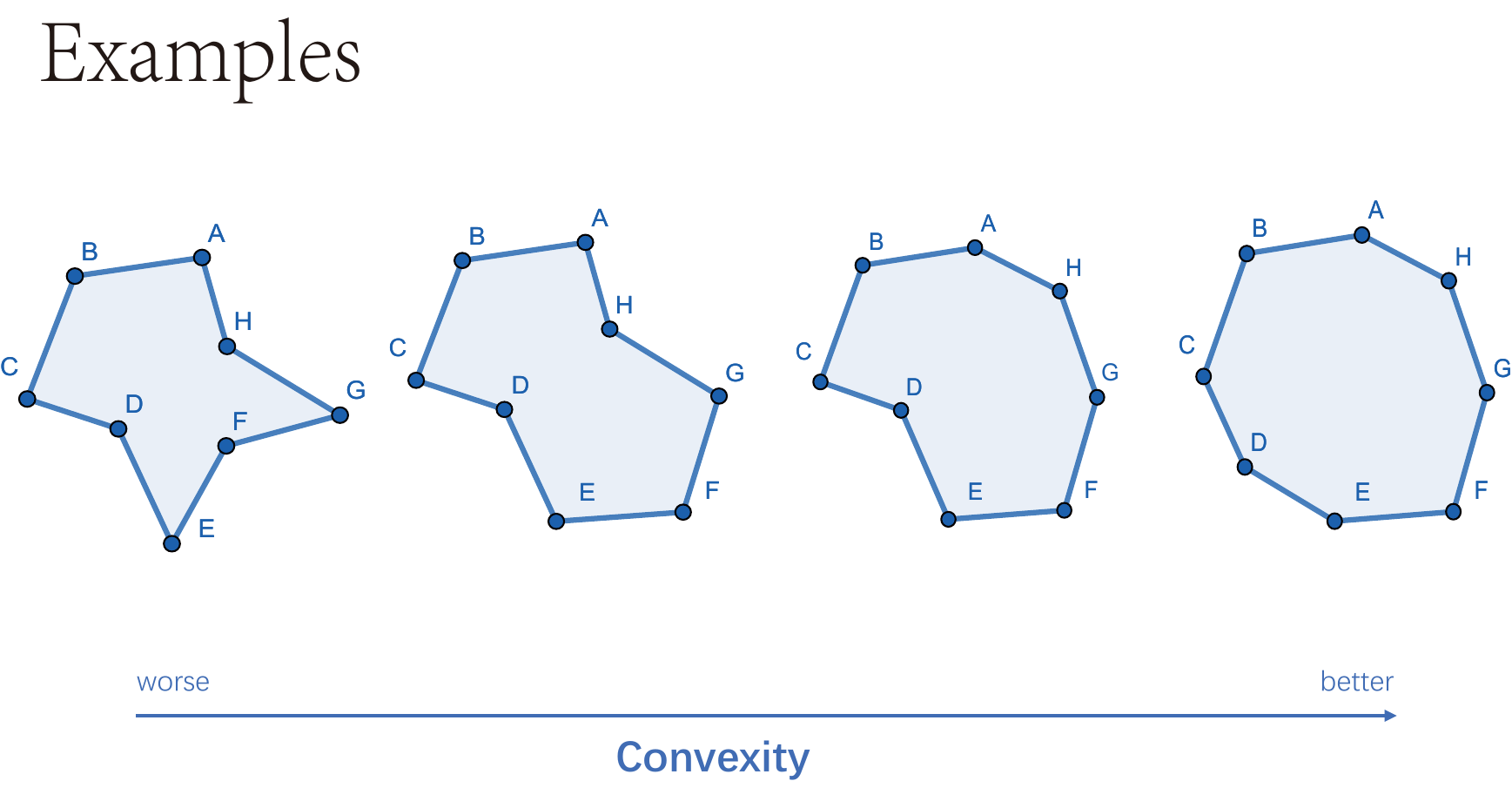}
\caption{Example of polygons with different convexity.}
\label{fig:tutorial4}
\end{figure}

\hspace*{\fill}\\
\noindent\textbf{What is convexity?}
Convexity describes how close a shape is to a convex shape.
For non-convex polygons, there are also polygons with better convexity or worse convexity.(\cref{fig:tutorial3})
Some examples are given to show polygons with different convexity. 
It can be seen that as the number and magnitude of depressions become smaller, the convexity of the polygon from left to right becomes better.
(\cref{fig:tutorial4})

\begin{figure}[!h]
\centering
\includegraphics[width=1\linewidth]{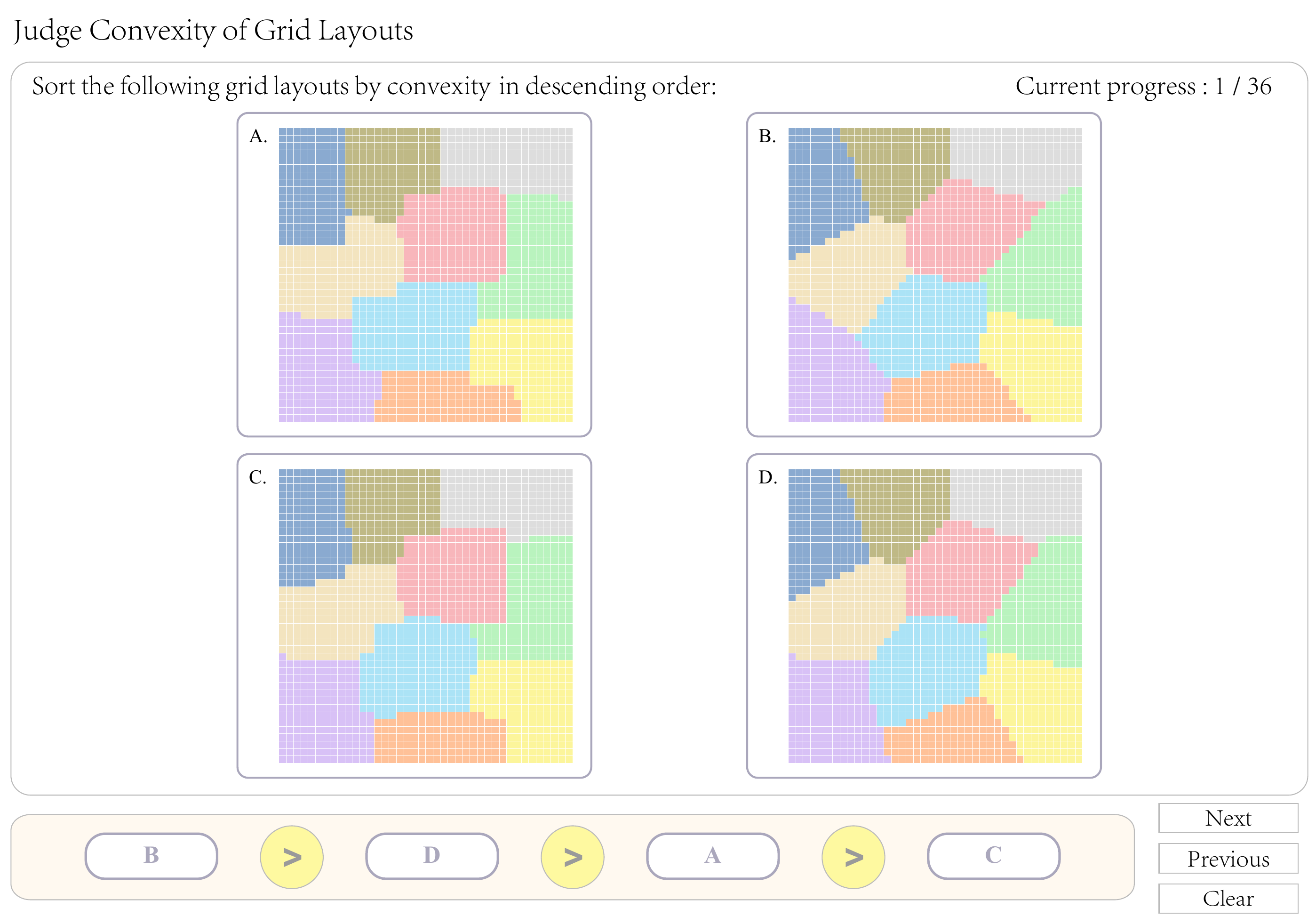}
\caption{Example interface of the user study.}
\label{fig:system}
\end{figure}

\hspace*{\fill}\\
\noindent\textbf{How to use the study system?}
The system of user study is introduced in the tutorial. (\cref{fig:system})
Four different grid visualizations are displayed in the system. 
Users need to click and sort grid visualizations according to their understanding of convexity. 
The sorting results will be displayed below. Users can drag and drop to modify the sorting results, or click ">" to modify it to "=" which indicate that the convexity of the visualizations on the left and right sides are similar.
After completing a question, the user can click the next button to proceed to the next question. If the user wants to modify the previous result, he can click the previous button to return to the previous question. We also provide a clear button to clear current answers.

\hspace{10cm}
\begin{figure*}[!htb]
\centering
\includegraphics[width=0.9\linewidth]{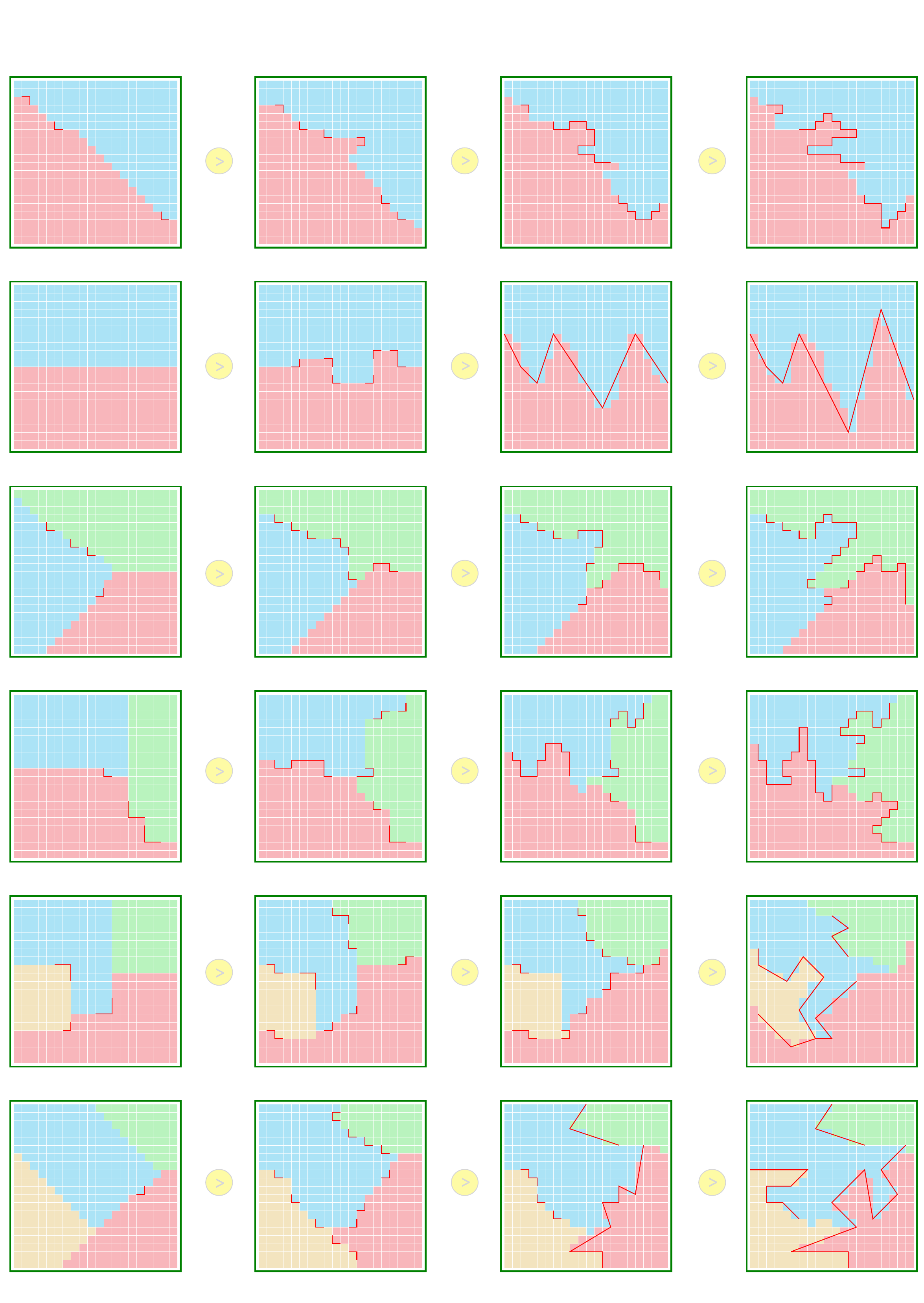}
\caption{Practice trials.}
\label{fig:practice}
\end{figure*}

\hspace*{\fill}\\
\noindent\textbf{Practice trials}
In the practice session, six practice trials were presented to participants to familiarize them with the concept of convexity and the use of the system.
After completing each exercise question, the system will check the answer and present the correct result of the question. 
At the same time, the positions that mainly affects the convexity of the visualization will be marked in red ink in the figure to help users understand the convexity.
 \cref{fig:practice} show these practice trials.

\newpage
\mbox{}
\newpage

\begin{figure*}[!hb]
\centering
\includegraphics[width=0.88\linewidth]{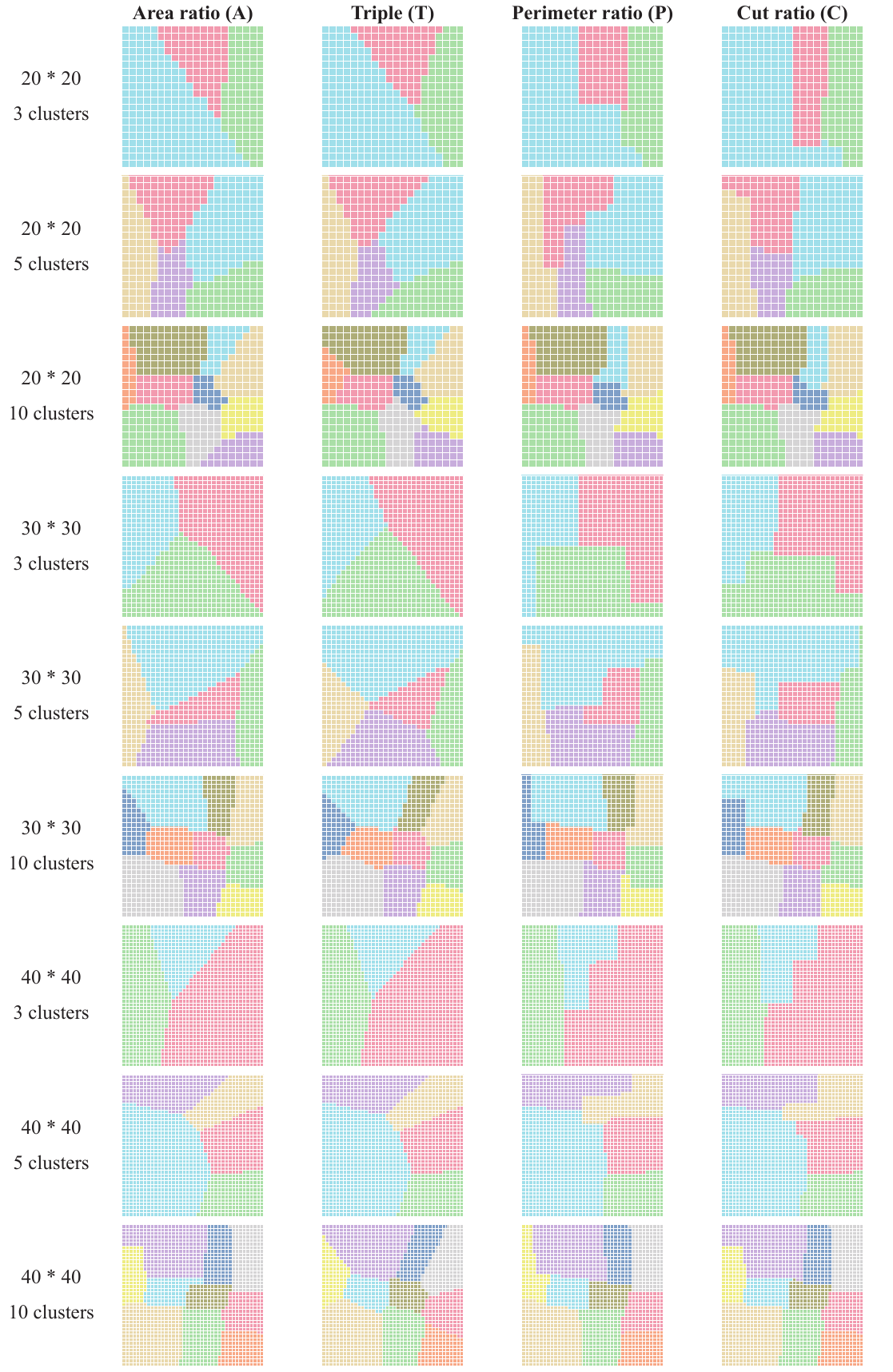}
\caption{Trials from dataset Animals~\cite{dataset-animals}.}
\label{fig:trials1}
\end{figure*}

\mbox{}
\newpage

\subsection{Formal study trials}

\begin{figure*}[!htb]
\centering
\includegraphics[width=0.88\linewidth]{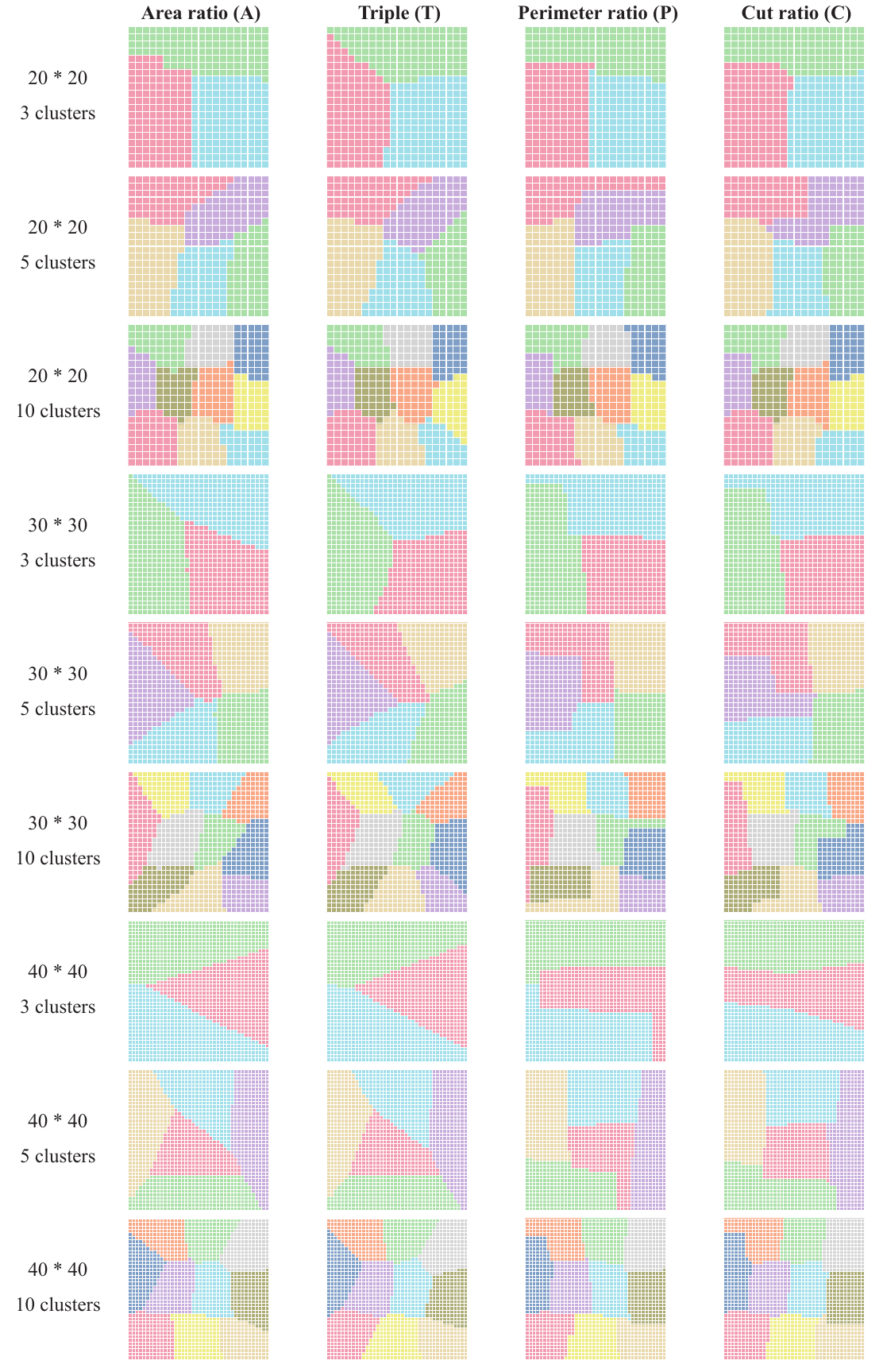}
\caption{Trials from dataset Cifar10~\cite{dataset-cifar10}.}
\label{fig:trials2}
\end{figure*}

\begin{figure*}[!htb]
\centering
\includegraphics[width=0.88\linewidth]{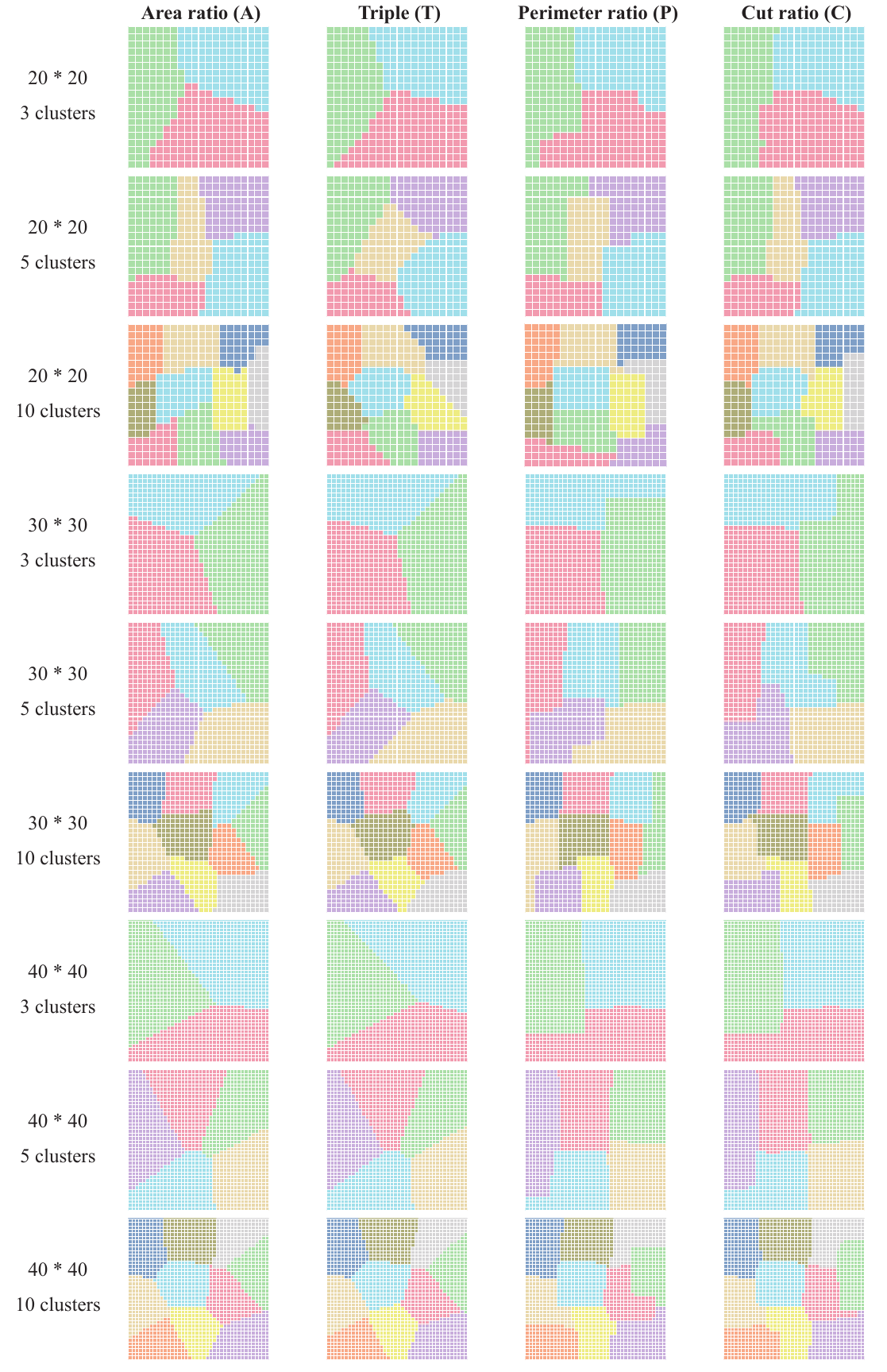}
\caption{Trials from dataset Mnist~\cite{dataset-mnist}.}
\label{fig:trials3}
\end{figure*}

\begin{figure*}[!htb]
\centering
\includegraphics[width=0.88\linewidth]{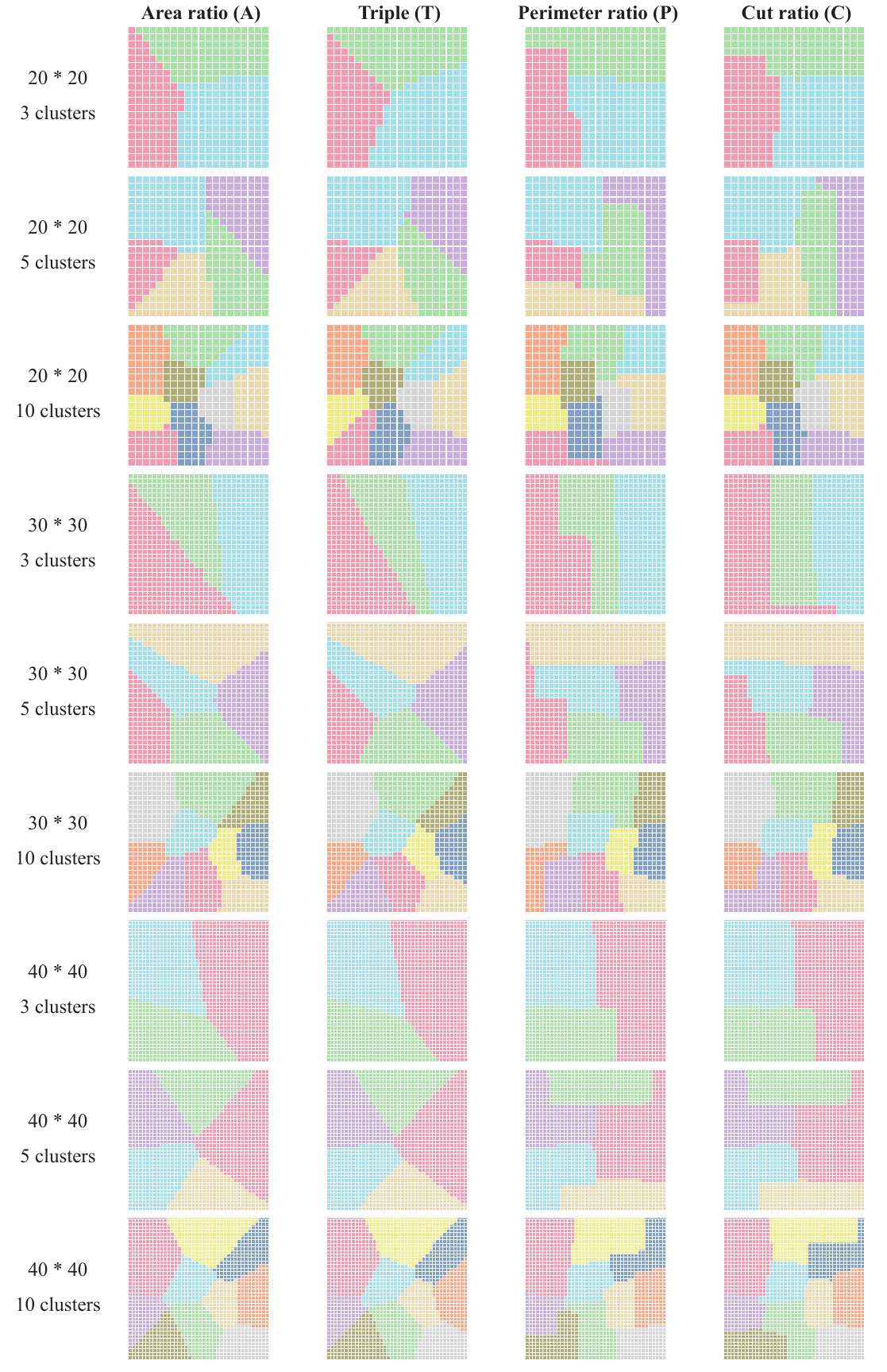}
\caption{Trials from dataset USPS~\cite{dataset-usps}}
\label{fig:trials4}
\end{figure*}

In the formal session, a total of 36 trials (3 grid sizes × 3 cluster numbers × 4 datasets) were evaluated.
\cref{fig:trials1} - \cref{fig:trials4} show these trials.


\clearpage
\subsection{Questionnaire}
\label{subsec:questionnaire}
Following the completion of all trials, participants were asked to fill out a questionnaire that 
included personal information and a question asking them to explain how 
they compare the convexity of different grid visualizations.
\\

\textbf{Part One: Basic Information}
\Qitem{ \Qq{Please select the range of your age.
}
\begin{Qlist}
\item 16 - 20 
\item 21 - 25
\item 26 - 30
\item 31 - 35
\item 36 - 40
\item 41 - 45
\item 46 - 50
\item 51 - 55
\item 56 - 60
\item More than 60
\end{Qlist}
}

\Qitem{ \Qq{Please select your gender. 
}
\begin{Qlist}
\item Male
\item Female
\end{Qlist}
}

\Qitem{ \Qq{Please select your education background.
}
\begin{Qlist}
\item High school and below
\item Bachelor
\item Master's degree
\item Doctoral degree
\end{Qlist}
}

\Qitem{ \Qq{Whether you have color blindness, color weakness or other diseases that affect visual judgment? }
\begin{Qlist}
\item No
\item Yes, illegible colors: \Qline{4cm}
\end{Qlist}
}

\Qitem{ \Qq{Please specify your contact information. \\ (telephone/email) } 

\vspace{2mm}
\Qline{7.2cm}}

\hspace*{\fill}\\
\indent\textbf{Part Two: Professional background}
\Qitem[]{ \Qq{ Are you familiar with the concepts of convexity and convex polygons? } 

Unfamiliar / Slightly familiar / Moderately familiar / Familiar / Very familiar

{Unfamiliar \Qrating{5} Very familiar}}

\Qitem[]{ \Qq{ Are you familiar with grid layouts? } 


{Unfamiliar \Qrating{5} Very familiar}}

\newpage
\textbf{Part Three: Open question}

\Qitem{ \Qq{How did you judge the convexity of graphics in the formal trials?}

For example, in the example below, among the factors such as the slope of the edge, the degree of curvature, the number of serrations, the number and size of the depressions, etc., which help you judge whether the grid layout
has better/worse convexity?

\begin{figure}[h]
\centering
\hspace*{5mm}
\includegraphics[width=0.8\linewidth]{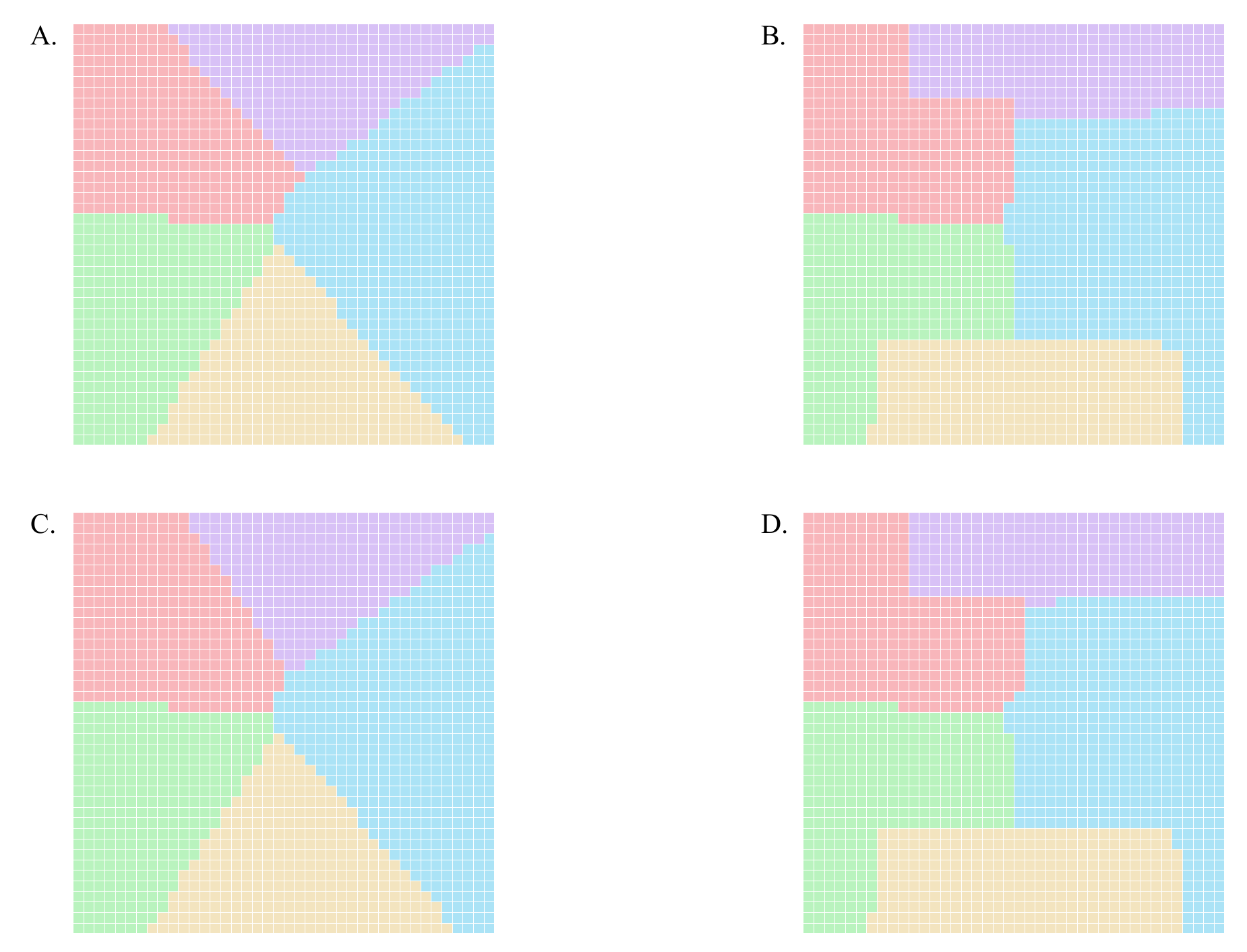}
\caption{Example}
\label{fig:interface}
\end{figure}

\vspace{2mm}
\Qline{7.5cm}

\Qline{7.5cm}

\Qline{7.5cm}
\vspace{2mm}
}



\newpage

\section{Experiments details}
\label{sec:exp-details-supple}

\subsection{Datasets}
\label{subsec:exp-datasets}
Ten datasets we used are from Xia~\etal's work ~\cite{xia2022interactive}.
They are Animals~\cite{dataset-animals}, CIFAR10~\cite{dataset-cifar10}, Indian Food~\cite{dataset-food}, Isolet~\cite{dataset-isolet}, MNIST~\cite{dataset-mnist}, Stanford Dogs~\cite{chen2020oodanalyzer}, Texture~\cite{dataset-texture}, USPS~\cite{dataset-usps}, Weather~\cite{dataset-weather} and Wifi~\cite{dataset-wifi}.
\\
We also used an additional dataset, OoD-Animals, which is from a real-world application~\cite{chen2020oodanalyzer}.
\\
It is about different images of different animals:  cat, dog, rabbit, wolf, and tiger.
The information of datasets are shown in \cref{table:dataset_information}.
\\

\begin{table}[!ht]
\setlength{\tabcolsep}{1.5em}
\centering
\caption{Datasets information.}
\begin{tabular}{lrrr}
\toprule
Dataset    & Size & Clusters & Type  \\
\midrule
Animals~\cite{dataset-animals} & 26179 & 10 & Image\\
CIFAR10~\cite{dataset-cifar10} & 60000 & 10 & Image\\
Indian Food~\cite{dataset-food} & 3625 & 11 & Image\\
Isolet~\cite{dataset-isolet} & 2352 & 8 & Text\\
MNIST~\cite{dataset-mnist} & 70000 & 10 & Image\\
Stanford Dogs~\cite{dataset-stanford-dogs} & 1291 & 7 & Image\\
Texture~\cite{dataset-texture} & 5500 & 11 & Text\\
USPS~\cite{dataset-usps} & 9298 & 10 & Image\\
Weather~\cite{dataset-weather} & 1156 & 4 & Image\\
Wifi~\cite{dataset-wifi} & 2000 & 4 & Tabular\\
OoD-Animals~\cite{chen2020oodanalyzer} & 26683 & 5 & Image\\
\bottomrule
\end{tabular}
\label{table:dataset_information}
\end{table}

\subsection{Pearson Correlations Between Convexity Measures}
\label{sec:exp-corr}
The correlation between convexity measures is calculated based on a set of diverse cluster shapes.
Therefore, the key is to generate a diverse set of cluster shapes that are similar to those that appear in a grid layout.
To achieve this, we generated multiple grid layouts using the baseline method and then extracted the shape of each cluster.
Specifically, we used the ten datasets from Xia~\etal's work ~\cite{xia2022interactive}, and generated 60 grid layouts using the baseline with each grid size (20x20, 30x30, 40x40).
Thus, we obtained 10x60x3=1800 grid layouts and then extracted corresponding cluster shapes.
If a cluster contained multiple disconnected components, we would only choose the largest connected one because those disconnected components usually have poor convexity at all measures, which cannot help evaluate the correlation between different measures.
In total, 9,689 different shapes are selected to evaluate the correlations between convexity measures.
\\




\subsection{Full Experiment Results in Evaluation}
\noindent\textbf{Layout generation}.
For each dataset, we began by sampling 20x20, 30x30, and 40x40 samples for each dataset.
We then generated t-SNE projections from these samples and used them as input for the baseline layout method.
However, because different rotations of the same t-SNE projection can produce different grid layouts, we rotated each projection with degrees $\pi/16*k,k=0,1,\ldots,7$.
To reduce the randomness in sampling, we repeated this entire process five times.
As a result, we generated a total of 120 layouts for each dataset (3 sizes x 8 rotations x 5 repetitions).

\noindent\textbf{Results}.
In the ablation study, \cref{table:ablation_proximity,table:ablation_compactness,table:ablation_area,table:ablation_triple,table:ablation_peri,table:ablation_cut} show the comparison of proximity, compactness, area ratio, triple ratio, perimeter ratio, and cut ratio of all the methods on 11 datasets.
The results are averaged over different grid sizes.
To demonstrate the effectiveness of our method,
\cref{table:full_20,table:full_30,table:full_40} show the comparison of proximity, compactness, area ratio, triple ratio, perimeter ratio, and cut ratio between baseline and Ours-T/Ours-P on 11 datasets with 3 different grid sizes.
\\

\newpage
\subsection{Examples}
\label{subsec:eva-example}
Here are examples of layouts generated by baseline method and our method, from different datasets.

\begin{figure}[!h]
\centering
\includegraphics[width=0.9\linewidth]{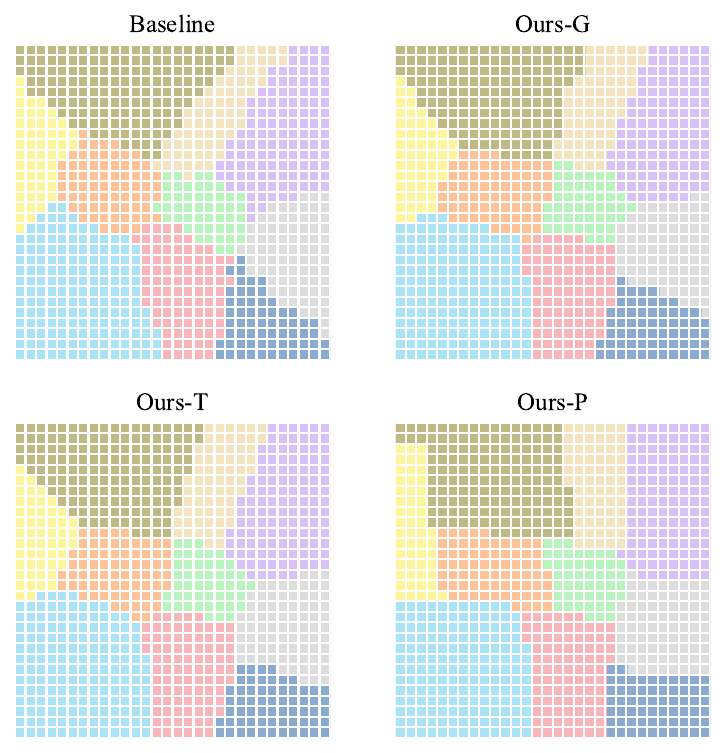}
\put(-132,240){\textbf{Animals}}
\label{fig:Animals}
\caption{}
\end{figure}

\begin{figure}[!h]
\centering
\includegraphics[width=0.9\linewidth]{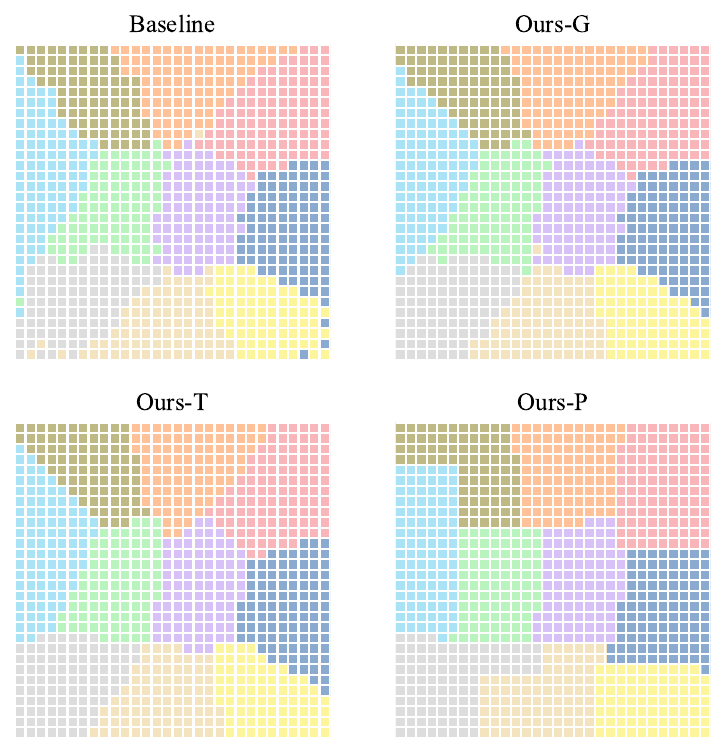}
\put(-132,240){\textbf{Cifar10}}
\label{fig:Cifar10}
\caption{}
\end{figure}

\newpage
\begin{figure}[!h]
\centering
\includegraphics[width=0.9\linewidth]{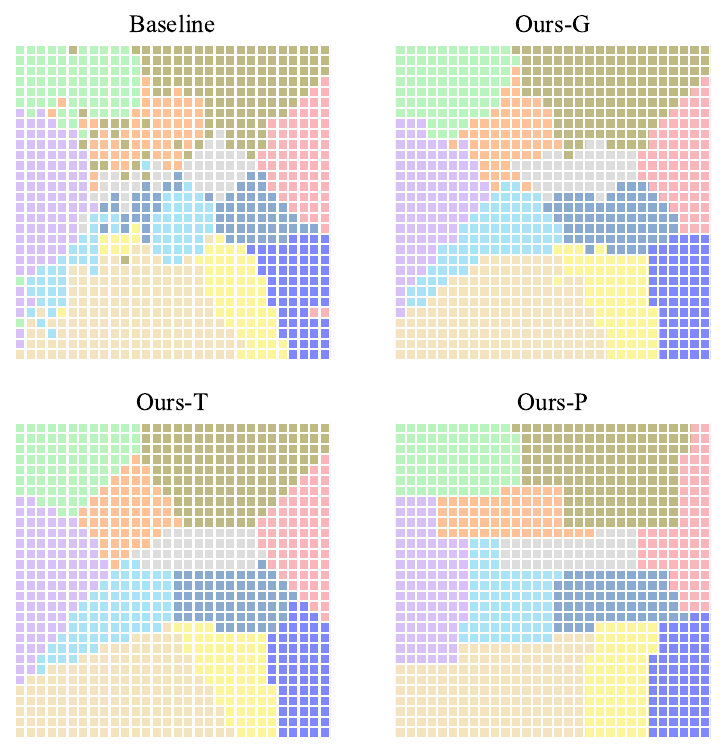}
\put(-138,240){\textbf{Indian Food}}
\label{fig:Indian Food}
\caption{}
\end{figure}

\begin{figure}[!h]
\centering
\includegraphics[width=0.9\linewidth]{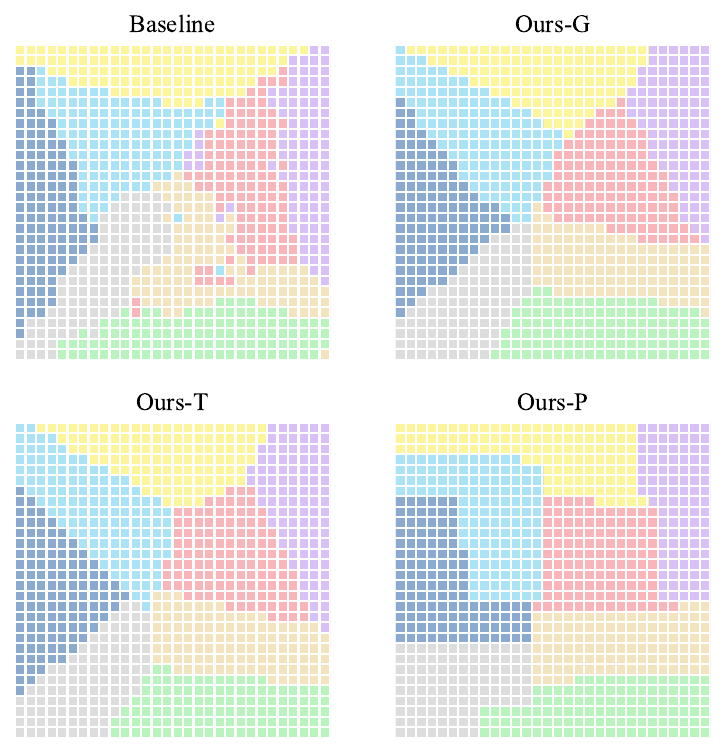}
\put(-126,240){\textbf{Isolet}}
\label{fig:Isolet}
\caption{}
\end{figure}

\newpage
\begin{figure}[!h]
\centering
\includegraphics[width=0.9\linewidth]{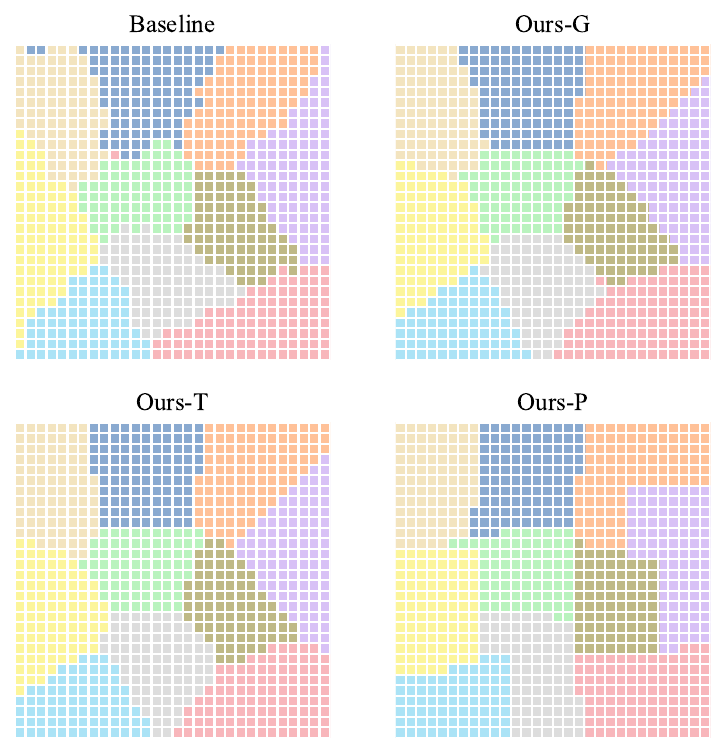}
\put(-128,240){\textbf{Mnist}}
\label{fig:Mnist}
\caption{}
\end{figure}

\begin{figure}[!h]
\centering
\includegraphics[width=0.9\linewidth]{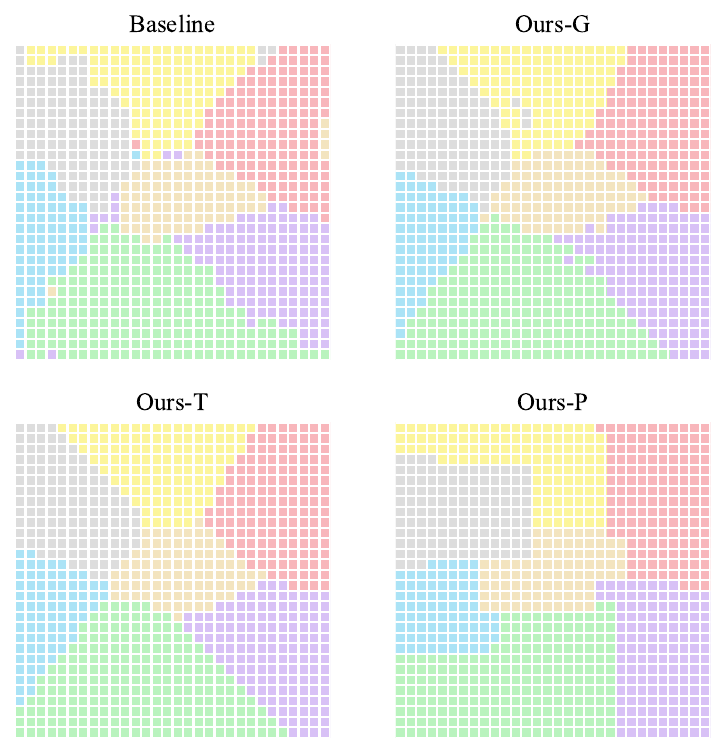}
\put(-143,240){\textbf{Stanford Dogs}}
\label{fig:Stanford Dogs}
\caption{}
\end{figure}

\newpage
\begin{figure}[!h]
\centering
\includegraphics[width=0.9\linewidth]{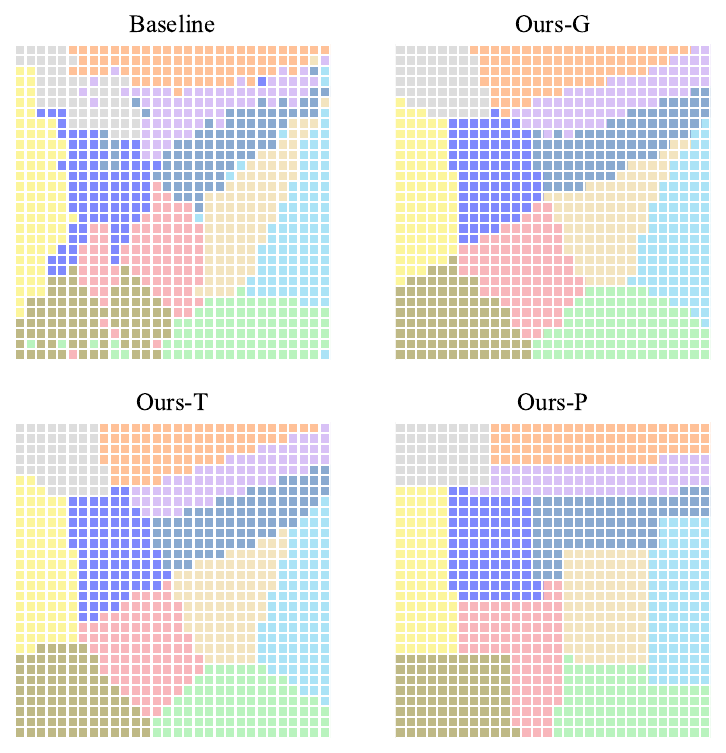}
\put(-130,240){\textbf{Texture}}
\label{fig:Texture}
\caption{}
\end{figure}

\begin{figure}[!h]
\centering
\includegraphics[width=0.9\linewidth]{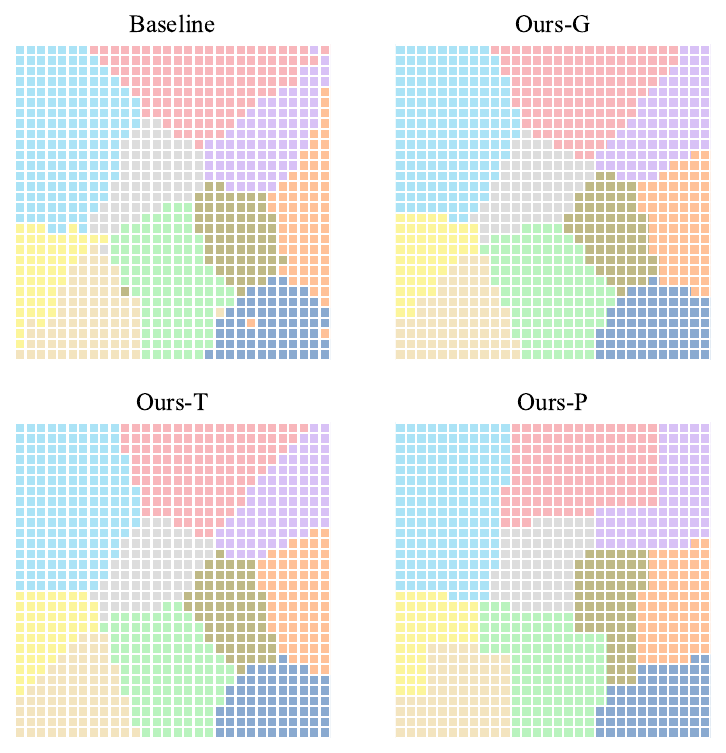}
\put(-128,240){\textbf{USPS}}
\label{fig:USPS}
\caption{}
\end{figure}

\newpage
\begin{figure}[!h]
\centering
\includegraphics[width=0.9\linewidth]{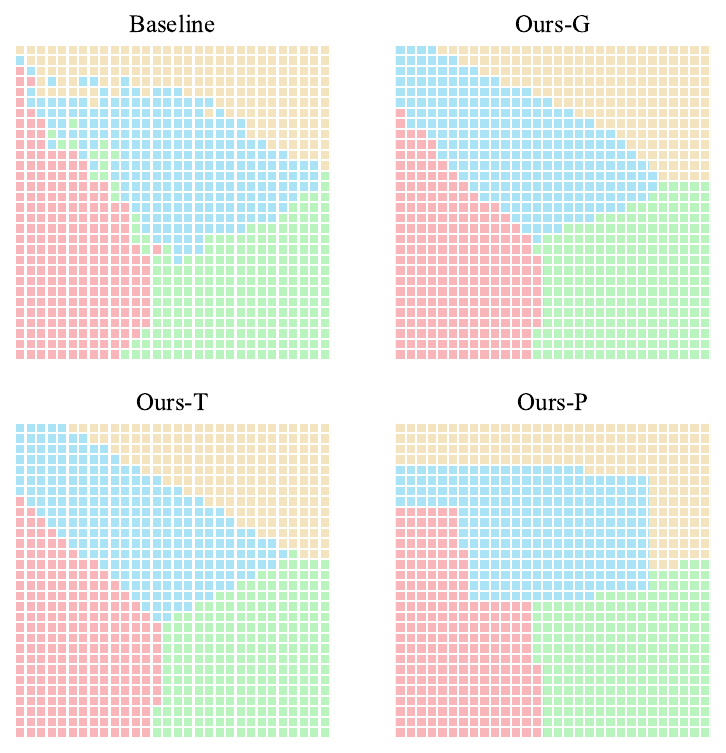}
\put(-133,240){\textbf{Weather}}
\label{fig:Weather}
\caption{}
\end{figure}

\begin{figure}[!h]
\centering
\includegraphics[width=0.9\linewidth]{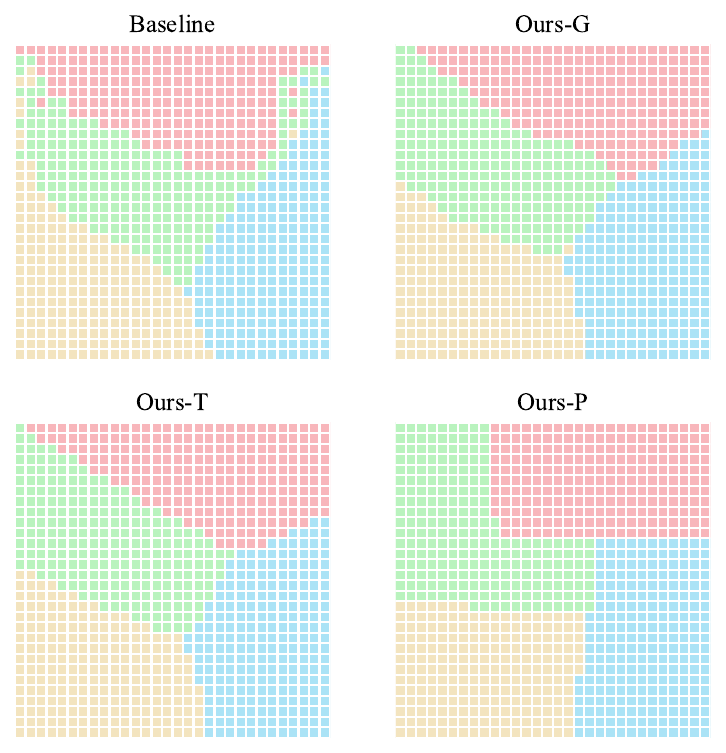}
\put(-126,240){\textbf{Wifi}}
\label{fig:Wifi}
\caption{}
\end{figure}

\newpage
\begin{figure}[!h]
\centering
\includegraphics[width=0.9\linewidth]{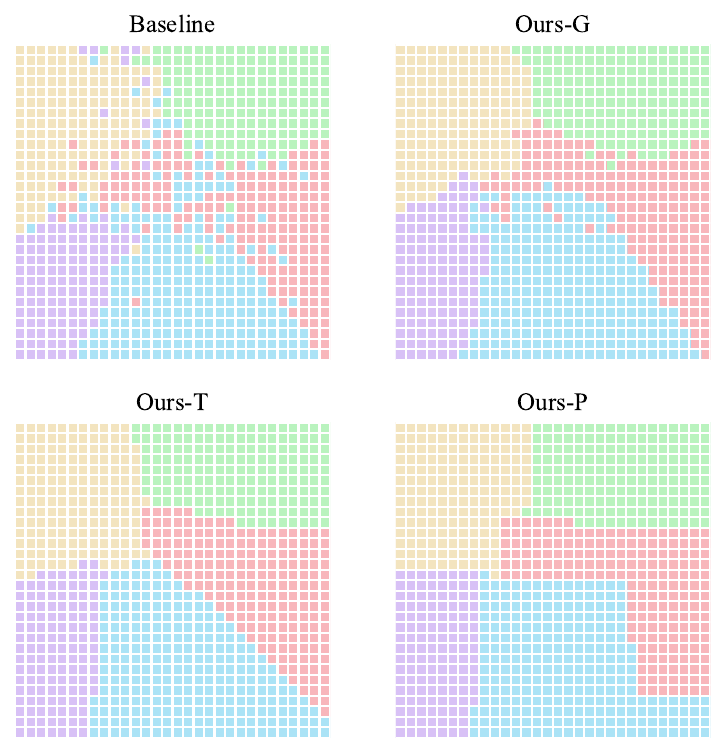}
\put(-143,240){\textbf{OoD-Animals}}
\label{fig:OoD-Animals}
\caption{}
\end{figure}

\begin{table*}[!t]
\setlength{\tabcolsep}{.9em}
\centering
\caption{Comparison of proximity of all the methods. G: global; L: local; T: triple ratio; P: perimeter ratio.}
\begin{tabular}{lrrrrrrrr}
\toprule
Dataset    & Baseline & Ours-G & Ours-L(T) & Ours-L(P) & Ours-L(T)-G & Ours-L(P)-G & Ours-G-L(T) & Ours-G-L(P)  \\
\midrule
Animals & \textbf{1.000} & 0.998 & 0.999 & 0.995 & 0.999 & 0.998 & 0.998 & 0.996\\
CIFAR10 & \textbf{1.000} & 0.999 & 0.999 & 0.996 & 0.999 & 0.999 & 0.999 & 0.997\\
Indian Food & \textbf{1.000} & 0.996 & 0.997 & 0.989 & 0.997 & 0.996 & 0.995 & 0.994\\
Isolet & \textbf{1.000} & 0.995 & 0.997 & 0.992 & 0.995 & 0.995 & 0.995 & 0.993\\
MNIST & \textbf{1.000} & 0.998 & 0.999 & 0.993 & 0.999 & 0.998 & 0.998 & 0.996\\
Stanford Dogs & \textbf{1.000} & 0.996 & 0.996 & 0.988 & 0.996 & 0.996 & 0.995 & 0.992\\
Texture & \textbf{1.000} & 0.996 & 0.998 & 0.990 & 0.996 & 0.996 & 0.995 & 0.994\\
USPS & \textbf{1.000} & 0.998 & 0.999 & 0.994 & 0.999 & 0.998 & 0.998 & 0.996\\
Weather & \textbf{1.000} & 0.992 & 0.998 & 0.991 & 0.997 & 0.992 & 0.991 & 0.987\\
Wifi & \textbf{1.000} & 0.991 & 0.998 & 0.994 & 0.992 & 0.991 & 0.991 & 0.989\\
OoD-Animals & \textbf{1.000} & 0.998 & 0.998 & 0.989 & 0.998 & 0.997 & 0.997 & 0.994\\
\midrule
Average & \textbf{1.000} & 0.996 & 0.998 & 0.992 & 0.997 & 0.996 & 0.996 & 0.994 \\
\bottomrule
\end{tabular}
\label{table:ablation_proximity}
\end{table*}

\begin{table*}[!t]
\setlength{\tabcolsep}{.9em}
\centering
\caption{Comparison of compactness of all the methods. G: global; L: local; T: triple ratio; P: perimeter ratio.}
\begin{tabular}{lrrrrrrrr}
\toprule
Dataset    & Baseline & Ours-G & Ours-L(T) & Ours-L(P) & Ours-L(T)-G & Ours-L(P)-G & Ours-G-L(T) & Ours-G-L(P)  \\
\midrule
Animals & 0.974 & \textbf{0.977} & 0.975 & 0.973 & 0.976 & \textbf{0.977} & \textbf{0.977} & 0.976\\
CIFAR10 & 0.980 & \textbf{0.981} & 0.980 & 0.978 & \textbf{0.981} & \textbf{0.981} & \textbf{0.981} & 0.980\\
Indian Food & 0.970 & \textbf{0.978} & 0.976 & 0.970 & 0.977 & \textbf{0.978} & \textbf{0.978} & 0.976\\
Isolet & 0.965 & \textbf{0.973} & 0.969 & 0.963 & \textbf{0.973} & \textbf{0.973} & \textbf{0.973} & 0.971\\
MNIST & 0.977 & \textbf{0.980} & 0.979 & 0.975 & \textbf{0.980} & \textbf{0.980} & \textbf{0.980} & 0.979\\
Stanford Dogs & 0.960 & \textbf{0.969} & 0.967 & 0.960 & \textbf{0.969} & \textbf{0.969} & \textbf{0.969} & 0.967\\
Texture & 0.973 & \textbf{0.979} & 0.976 & 0.970 & \textbf{0.979} & \textbf{0.979} & \textbf{0.979} & 0.977\\
USPS & 0.976 & \textbf{0.979} & 0.977 & 0.974 & 0.978 & \textbf{0.979} & \textbf{0.979} & 0.978\\
Weather & 0.936 & \textbf{0.944} & 0.939 & 0.937 & 0.940 & 0.942 & \textbf{0.944} & 0.941\\
Wifi & 0.939 & \textbf{0.950} & 0.942 & 0.937 & 0.949 & 0.949 & \textbf{0.950} & 0.948\\
OoD-Animals & 0.953 & \textbf{0.960} & 0.959 & 0.952 & \textbf{0.960} & \textbf{0.960} & \textbf{0.960} & 0.958\\
\midrule
Average & 0.964 & \textbf{0.970} & 0.967 & 0.963 & 0.969 & \textbf{0.970} & \textbf{0.970} & 0.968 \\
\bottomrule
\end{tabular}
\label{table:ablation_compactness}
\end{table*}

\begin{table*}[!t]
\setlength{\tabcolsep}{.9em}
\centering
\caption{Comparison of area ratio of all the methods. G: global; L: local; T: triple ratio; P: perimeter ratio.}
\begin{tabular}{lrrrrrrrr}
\toprule
Dataset    & Baseline & Ours-G & Ours-L(T) & Ours-L(P) & Ours-L(T)-G & Ours-L(P)-G & Ours-G-L(T) & Ours-G-L(P)  \\
\midrule
Animals & 0.775 & 0.893 & 0.897 & 0.872 & 0.884 & 0.896 & \textbf{0.910} & 0.900\\
CIFAR10 & 0.786 & 0.893 & 0.899 & 0.869 & 0.885 & 0.893 & \textbf{0.910} & 0.901\\
Indian Food & 0.555 & 0.846 & 0.889 & 0.818 & 0.832 & 0.840 & \textbf{0.901} & 0.885\\
Isolet & 0.593 & 0.867 & 0.886 & 0.827 & 0.855 & 0.870 & \textbf{0.905} & 0.891\\
MNIST & 0.718 & 0.886 & 0.899 & 0.868 & 0.862 & 0.884 & \textbf{0.908} & 0.901\\
Stanford Dogs & 0.577 & 0.860 & 0.907 & 0.846 & 0.851 & 0.858 & \textbf{0.919} & 0.890\\
Texture & 0.617 & 0.856 & 0.877 & 0.819 & 0.851 & 0.858 & \textbf{0.889} & 0.870\\
USPS & 0.739 & 0.892 & 0.896 & 0.875 & 0.877 & 0.891 & \textbf{0.910} & 0.904\\
Weather & 0.738 & 0.903 & 0.906 & 0.872 & 0.874 & 0.918 & \textbf{0.926} & 0.877\\
Wifi & 0.724 & 0.921 & 0.916 & 0.845 & 0.915 & 0.921 & \textbf{0.936} & 0.916\\
OoD-Animals & 0.534 & 0.889 & 0.924 & 0.862 & 0.877 & 0.886 & \textbf{0.932} & 0.914\\
\midrule
Average & 0.669 & 0.882 & 0.900 & 0.852 & 0.869 & 0.883 & \textbf{0.913} & 0.895 \\

\bottomrule
\end{tabular}
\label{table:ablation_area}
\end{table*}

\begin{table*}[!t]
\setlength{\tabcolsep}{.9em}
\centering
\caption{Comparison of triple ratio of all the methods. G: global; L: local; T: triple ratio; P: perimeter ratio.}
\begin{tabular}{lrrrrrrrr}
\toprule
Dataset    & Baseline & Ours-G & Ours-L(T) & Ours-L(P) & Ours-L(T)-G & Ours-L(P)-G & Ours-G-L(T) & Ours-G-L(P)  \\
\midrule
Animals & 0.977 & 0.995 & 0.996 & 0.973 & 0.994 & 0.995 & \textbf{0.997} & 0.983\\
CIFAR10 & 0.975 & 0.994 & 0.996 & 0.961 & 0.993 & 0.994 & \textbf{0.997} & 0.979\\
Indian Food & 0.889 & 0.986 & 0.996 & 0.947 & 0.985 & 0.984 & \textbf{0.997} & 0.981\\
Isolet & 0.893 & 0.986 & 0.993 & 0.936 & 0.983 & 0.987 & \textbf{0.996} & 0.974\\
MNIST & 0.963 & 0.993 & 0.996 & 0.964 & 0.989 & 0.991 & \textbf{0.997} & 0.982\\
Stanford Dogs & 0.909 & 0.984 & 0.997 & 0.949 & 0.983 & 0.983 & \textbf{0.998} & 0.976\\
Texture & 0.893 & 0.985 & 0.991 & 0.929 & 0.983 & 0.985 & \textbf{0.994} & 0.964\\
USPS & 0.968 & 0.995 & 0.996 & 0.971 & 0.992 & 0.994 & \textbf{0.997} & 0.985\\
Weather & 0.952 & 0.993 & 0.995 & 0.963 & 0.989 & 0.996 & \textbf{0.997} & 0.965\\
Wifi & 0.953 & 0.997 & 0.996 & 0.944 & 0.996 & 0.997 & \textbf{0.998} & 0.982\\
OoD-Animals & 0.925 & 0.993 & \textbf{0.998} & 0.961 & 0.993 & 0.993 & \textbf{0.998} & 0.984\\
\midrule
Average & 0.936 & 0.991 & 0.995 & 0.954 & 0.989 & 0.991 & \textbf{0.997} & 0.978 \\
\bottomrule
\end{tabular}
\label{table:ablation_triple}
\end{table*}

\begin{table*}[!t]
\setlength{\tabcolsep}{.9em}
\centering
\caption{Comparison of perimeter ratio of all the methods. G: global; L: local; T: triple ratio; P: perimeter ratio.}
\begin{tabular}{lrrrrrrrr}
\toprule
Dataset    & Baseline & Ours-G & Ours-L(T) & Ours-L(P) & Ours-L(T)-G & Ours-L(P)-G & Ours-G-L(T) & Ours-G-L(P)  \\
\midrule
Animals & 0.830 & 0.859 & 0.853 & 0.926 & 0.851 & 0.861 & 0.872 & \textbf{0.935}\\
CIFAR10 & 0.834 & 0.855 & 0.850 & 0.925 & 0.850 & 0.853 & 0.868 & \textbf{0.934}\\
Indian Food & 0.807 & 0.804 & 0.854 & 0.916 & 0.809 & 0.800 & 0.861 & \textbf{0.929}\\
Isolet & 0.784 & 0.818 & 0.858 & 0.921 & 0.818 & 0.824 & 0.864 & \textbf{0.936}\\
MNIST & 0.852 & 0.845 & 0.858 & 0.926 & 0.844 & 0.844 & 0.864 & \textbf{0.933}\\
Stanford Dogs & 0.868 & 0.773 & 0.854 & 0.926 & 0.777 & 0.772 & 0.868 & \textbf{0.933}\\
Texture & 0.794 & 0.828 & 0.857 & 0.922 & 0.828 & 0.833 & 0.850 & \textbf{0.927}\\
USPS & 0.861 & 0.851 & 0.854 & 0.930 & 0.846 & 0.850 & 0.865 & \textbf{0.933}\\
Weather & 0.790 & 0.858 & 0.863 & \textbf{0.940} & 0.849 & 0.880 & 0.864 & 0.937\\
Wifi & 0.784 & 0.867 & 0.861 & 0.929 & 0.867 & 0.867 & 0.875 & \textbf{0.948}\\
OoD-Animals & 0.724 & 0.812 & 0.862 & 0.929 & 0.825 & 0.807 & 0.873 & \textbf{0.944}\\
\midrule
Average & 0.812 & 0.834 & 0.857 & 0.926 & 0.833 & 0.835 & 0.866 & \textbf{0.935} \\

\bottomrule
\end{tabular}
\label{table:ablation_peri}
\end{table*}

\begin{table*}[!t]
\setlength{\tabcolsep}{.9em}
\centering
\caption{Comparison of cut ratio of all the methods. G: global; L: local; T: triple ratio; P: perimeter ratio.}
\begin{tabular}{lrrrrrrrr}
\toprule
Dataset    & Baseline & Ours-G & Ours-L(T) & Ours-L(P) & Ours-L(T)-G & Ours-L(P)-G & Ours-G-L(T) & Ours-G-L(P)  \\
\midrule
Animals & 0.838 & 0.887 & 0.872 & 0.920 & 0.879 & 0.888 & 0.895 & \textbf{0.936}\\
CIFAR10 & 0.843 & 0.890 & 0.876 & 0.920 & 0.883 & 0.889 & 0.899 & \textbf{0.937}\\
Indian Food & 0.763 & 0.849 & 0.867 & 0.893 & 0.846 & 0.844 & 0.885 & \textbf{0.925}\\
Isolet & 0.776 & 0.858 & 0.868 & 0.901 & 0.855 & 0.861 & 0.885 & \textbf{0.932}\\
MNIST & 0.848 & 0.887 & 0.889 & 0.918 & 0.882 & 0.886 & 0.900 & \textbf{0.936}\\
Stanford Dogs & 0.786 & 0.836 & 0.872 & 0.912 & 0.835 & 0.835 & 0.893 & \textbf{0.931}\\
Texture & 0.781 & 0.856 & 0.860 & 0.898 & 0.855 & 0.860 & 0.870 & \textbf{0.921}\\
USPS & 0.839 & 0.888 & 0.878 & 0.922 & 0.879 & 0.887 & 0.898 & \textbf{0.937}\\
Weather & 0.803 & 0.862 & 0.856 & \textbf{0.927} & 0.852 & 0.889 & 0.868 & \textbf{0.927}\\
Wifi & 0.800 & 0.890 & 0.866 & 0.910 & 0.889 & 0.890 & 0.896 & \textbf{0.947}\\
OoD-Animals & 0.744 & 0.868 & 0.892 & 0.920 & 0.870 & 0.865 & 0.904 & \textbf{0.946}\\
\midrule
Average & 0.802 & 0.870 & 0.872 & 0.913 & 0.866 & 0.872 & 0.890 & \textbf{0.934} \\

\bottomrule
\end{tabular}
\label{table:ablation_cut}
\end{table*}

\begin{table*}[!t]
\fontsize{8}{8}\selectfont
\setlength{\tabcolsep}{.24em}
\centering
\caption{Comparison of four convexity measures on 11 datasets with grid size of 20x20.}
\begin{tabular}{l|ccc|ccc|ccc|ccc|ccc|ccc}
\toprule
\multirow{2}{*}{Dataset} & \multicolumn{3}{c|}{Proximity}  & \multicolumn{3}{c|}{Compactness} & \multicolumn{3}{c|}{Area ratio}  & \multicolumn{3}{c|}{Triple ratio} & \multicolumn{3}{c|}{Perimeter ratio} & \multicolumn{3}{c}{Cut ratio} \\
& Basel. & Ours-T & Ours-P & Basel. & Ours-T & Ours-P & Basel. & Ours-T & Ours-P & Basel. & Ours-T & Ours-P & Basel. & Ours-T & Ours-P & Basel. & Ours-T & Ours-P \\
\midrule
Animals & \textbf{1.000} & 0.998 & 0.995 & 0.975 & \textbf{0.977} & 0.975 & 0.833 & \textbf{0.893} & \textbf{0.893} & 0.981 & \textbf{0.995} & 0.980 & 0.840 & 0.878 & \textbf{0.935} & 0.860 & 0.896 & \textbf{0.932}\\
CIFAR10 & \textbf{1.000} & 0.999 & 0.996 & 0.980 & \textbf{0.981} & 0.979 & 0.831 & \textbf{0.892} & 0.891 & 0.979 & \textbf{0.995} & 0.973 & 0.832 & 0.873 & \textbf{0.934} & 0.859 & 0.897 & \textbf{0.932}\\
Indian Food & \textbf{1.000} & 0.997 & 0.995 & 0.975 & \textbf{0.979} & 0.977 & 0.720 & 0.893 & \textbf{0.898} & 0.944 & \textbf{0.996} & 0.983 & 0.843 & 0.874 & \textbf{0.936} & 0.831 & 0.899 & \textbf{0.934}\\
Isolet & \textbf{1.000} & 0.995 & 0.994 & 0.968 & \textbf{0.974} & 0.972 & 0.696 & \textbf{0.893} & 0.888 & 0.920 & \textbf{0.994} & 0.973 & 0.838 & 0.874 & \textbf{0.935} & 0.826 & 0.894 & \textbf{0.931}\\
MNIST & \textbf{1.000} & 0.998 & 0.995 & 0.977 & \textbf{0.980} & 0.978 & 0.746 & \textbf{0.891} & \textbf{0.891} & 0.957 & \textbf{0.995} & 0.975 & 0.863 & 0.873 & \textbf{0.934} & 0.860 & 0.901 & \textbf{0.932}\\
Stanford Dogs & \textbf{1.000} & 0.995 & 0.991 & 0.961 & \textbf{0.969} & 0.967 & 0.644 & \textbf{0.907} & 0.895 & 0.912 & \textbf{0.997} & 0.978 & 0.898 & 0.876 & \textbf{0.937} & 0.819 & 0.901 & \textbf{0.934}\\
Texture & \textbf{1.000} & 0.995 & 0.994 & 0.974 & \textbf{0.980} & 0.977 & 0.733 & \textbf{0.869} & 0.865 & 0.925 & \textbf{0.990} & 0.956 & 0.848 & 0.857 & \textbf{0.929} & 0.837 & 0.874 & \textbf{0.920}\\
USPS & \textbf{1.000} & 0.997 & 0.995 & 0.976 & \textbf{0.979} & 0.978 & 0.777 & \textbf{0.895} & \textbf{0.895} & 0.969 & \textbf{0.995} & 0.981 & 0.867 & 0.876 & \textbf{0.932} & 0.860 & 0.902 & \textbf{0.933}\\
Weather & \textbf{1.000} & 0.995 & 0.988 & 0.939 & \textbf{0.942} & 0.940 & 0.824 & \textbf{0.909} & 0.907 & 0.973 & \textbf{0.995} & 0.977 & 0.844 & 0.877 & \textbf{0.949} & 0.847 & 0.878 & \textbf{0.944}\\
Wifi & \textbf{1.000} & 0.991 & 0.989 & 0.942 & \textbf{0.950} & 0.948 & 0.824 & \textbf{0.924} & 0.912 & 0.974 & \textbf{0.997} & 0.979 & 0.834 & 0.884 & \textbf{0.947} & 0.844 & 0.901 & \textbf{0.945}\\
OoD-Animals & \textbf{1.000} & 0.997 & 0.993 & 0.953 & \textbf{0.960} & 0.958 & 0.624 & \textbf{0.918} & 0.915 & 0.922 & \textbf{0.997} & 0.985 & 0.767 & 0.879 & \textbf{0.944} & 0.781 & 0.907 & \textbf{0.947}\\

\midrule
Average & \textbf{1.000} & 0.996 & 0.993 & 0.965 & \textbf{0.970} & 0.968 & 0.750 & \textbf{0.898} & 0.896 & 0.951 & \textbf{0.995} & 0.976 & 0.843 & 0.875 & \textbf{0.938} & 0.839 & 0.895 & \textbf{0.935} \\
\bottomrule
\end{tabular}
\label{table:full_20}
\end{table*}

\begin{table*}[!t]
\fontsize{8}{8}\selectfont
\setlength{\tabcolsep}{.24em}
\centering
\caption{Comparison of four convexity measures on 11 datasets with grid size of 30x30.}

\begin{tabular}{l|ccc|ccc|ccc|ccc|ccc|ccc}
\toprule
\multirow{2}{*}{Dataset} & \multicolumn{3}{c|}{Proximity}  & \multicolumn{3}{c|}{Compactness} & \multicolumn{3}{c|}{Area ratio}  & \multicolumn{3}{c|}{Triple ratio} & \multicolumn{3}{c|}{Perimeter ratio} & \multicolumn{3}{c}{Cut ratio} \\
& Basel. & Ours-T & Ours-P & Basel. & Ours-T & Ours-P & Basel. & Ours-T & Ours-P & Basel. & Ours-T & Ours-P & Basel. & Ours-T & Ours-P & Basel. & Ours-T & Ours-P \\
\midrule
Animals & \textbf{1.000} & 0.998 & 0.996 & 0.974 & \textbf{0.977} & \textbf{0.977} & 0.767 & \textbf{0.912} & 0.900 & 0.976 & \textbf{0.997} & 0.983 & 0.834 & 0.871 & \textbf{0.934} & 0.838 & 0.896 & \textbf{0.936}\\
CIFAR10 & \textbf{1.000} & 0.999 & 0.997 & 0.980 & \textbf{0.981} & \textbf{0.981} & 0.799 & \textbf{0.914} & 0.904 & 0.978 & \textbf{0.997} & 0.981 & 0.839 & 0.870 & \textbf{0.934} & 0.849 & 0.901 & \textbf{0.939}\\
Indian Food & \textbf{1.000} & 0.996 & 0.995 & 0.971 & \textbf{0.978} & 0.977 & 0.542 & \textbf{0.905} & 0.892 & 0.893 & \textbf{0.997} & 0.985 & 0.808 & 0.861 & \textbf{0.929} & 0.758 & 0.887 & \textbf{0.927}\\
Isolet & \textbf{1.000} & 0.995 & 0.993 & 0.965 & \textbf{0.973} & 0.971 & 0.592 & \textbf{0.906} & 0.892 & 0.888 & \textbf{0.996} & 0.975 & 0.775 & 0.861 & \textbf{0.936} & 0.772 & 0.883 & \textbf{0.932}\\
MNIST & \textbf{1.000} & 0.998 & 0.997 & 0.978 & \textbf{0.981} & 0.980 & 0.738 & \textbf{0.911} & 0.907 & 0.973 & \textbf{0.997} & 0.985 & 0.856 & 0.865 & \textbf{0.934} & 0.857 & 0.902 & \textbf{0.939}\\
Stanford Dogs & \textbf{1.000} & 0.995 & 0.993 & 0.960 & \textbf{0.969} & 0.967 & 0.557 & \textbf{0.922} & 0.889 & 0.911 & \textbf{0.998} & 0.975 & 0.889 & 0.868 & \textbf{0.931} & 0.787 & 0.895 & \textbf{0.930}\\
Texture & \textbf{1.000} & 0.996 & 0.995 & 0.973 & \textbf{0.980} & 0.978 & 0.636 & \textbf{0.893} & 0.877 & 0.908 & \textbf{0.995} & 0.969 & 0.805 & 0.848 & \textbf{0.928} & 0.784 & 0.871 & \textbf{0.923}\\
USPS & \textbf{1.000} & 0.998 & 0.997 & 0.976 & \textbf{0.979} & 0.978 & 0.758 & \textbf{0.911} & 0.905 & 0.975 & \textbf{0.998} & 0.986 & 0.864 & 0.863 & \textbf{0.933} & 0.845 & 0.898 & \textbf{0.937}\\
Weather & \textbf{1.000} & 0.990 & 0.987 & 0.937 & \textbf{0.945} & 0.942 & 0.726 & \textbf{0.930} & 0.872 & 0.951 & \textbf{0.997} & 0.961 & 0.787 & 0.861 & \textbf{0.936} & 0.797 & 0.865 & \textbf{0.924}\\
Wifi & \textbf{1.000} & 0.991 & 0.988 & 0.939 & \textbf{0.950} & 0.948 & 0.702 & \textbf{0.939} & 0.921 & 0.952 & \textbf{0.998} & 0.986 & 0.789 & 0.875 & \textbf{0.949} & 0.795 & 0.898 & \textbf{0.950}\\
OoD-Animals & \textbf{1.000} & 0.997 & 0.994 & 0.953 & \textbf{0.960} & 0.959 & 0.488 & \textbf{0.933} & 0.914 & 0.919 & \textbf{0.998} & 0.984 & 0.731 & 0.871 & \textbf{0.944} & 0.734 & 0.903 & \textbf{0.945}\\

\midrule
Average & \textbf{1.000} & 0.996 & 0.994 & 0.964 & \textbf{0.970} & 0.969 & 0.664 & \textbf{0.916} & 0.897 & 0.938 & \textbf{0.997} & 0.979 & 0.816 & 0.865 & \textbf{0.935} & 0.801 & 0.891 & \textbf{0.935} \\
\bottomrule
\end{tabular}
\label{table:full_30}
\end{table*}

\begin{table*}[!t]
\fontsize{8}{8}\selectfont
\setlength{\tabcolsep}{.24em}
\centering
\caption{Comparison of four convexity measures on 11 datasets with grid size of 40x40.}
\begin{tabular}{l|ccc|ccc|ccc|ccc|ccc|ccc}
\toprule
\multirow{2}{*}{Dataset} & \multicolumn{3}{c|}{Proximity}  & \multicolumn{3}{c|}{Compactness} & \multicolumn{3}{c|}{Area ratio}  & \multicolumn{3}{c|}{Triple ratio} & \multicolumn{3}{c|}{Perimeter ratio} & \multicolumn{3}{c}{Cut ratio} \\
& Basel. & Ours-T & Ours-P & Basel. & Ours-T & Ours-P & Basel. & Ours-T & Ours-P & Basel. & Ours-T & Ours-P & Basel. & Ours-T & Ours-P & Basel. & Ours-T & Ours-P \\
\midrule
Animals & \textbf{1.000} & 0.998 & 0.996 & 0.974 & \textbf{0.977} & \textbf{0.977} & 0.726 & \textbf{0.925} & 0.906 & 0.975 & \textbf{0.998} & 0.986 & 0.816 & 0.866 & \textbf{0.935} & 0.816 & 0.894 & \textbf{0.939}\\
CIFAR10 & \textbf{1.000} & 0.999 & 0.998 & 0.979 & \textbf{0.981} & \textbf{0.981} & 0.727 & \textbf{0.924} & 0.907 & 0.969 & \textbf{0.998} & 0.984 & 0.830 & 0.861 & \textbf{0.935} & 0.820 & 0.898 & \textbf{0.940}\\
Indian Food & \textbf{1.000} & 0.993 & 0.992 & 0.965 & \textbf{0.977} & 0.975 & 0.403 & \textbf{0.906} & 0.864 & 0.830 & \textbf{0.998} & 0.976 & 0.769 & 0.848 & \textbf{0.922} & 0.701 & 0.868 & \textbf{0.913}\\
Isolet & \textbf{1.000} & 0.994 & 0.993 & 0.963 & \textbf{0.973} & 0.971 & 0.491 & \textbf{0.917} & 0.892 & 0.870 & \textbf{0.997} & 0.974 & 0.740 & 0.855 & \textbf{0.938} & 0.731 & 0.877 & \textbf{0.932}\\
MNIST & \textbf{1.000} & 0.998 & 0.997 & 0.977 & \textbf{0.981} & 0.980 & 0.670 & \textbf{0.922} & 0.905 & 0.960 & \textbf{0.998} & 0.985 & 0.838 & 0.855 & \textbf{0.931} & 0.828 & 0.896 & \textbf{0.938}\\
Stanford Dogs & \textbf{1.000} & 0.995 & 0.993 & 0.960 & \textbf{0.969} & 0.968 & 0.529 & \textbf{0.928} & 0.886 & 0.904 & \textbf{0.998} & 0.977 & 0.816 & 0.858 & \textbf{0.931} & 0.753 & 0.884 & \textbf{0.929}\\
Texture & \textbf{1.000} & 0.995 & 0.994 & 0.970 & \textbf{0.979} & 0.977 & 0.481 & \textbf{0.905} & 0.869 & 0.845 & \textbf{0.996} & 0.966 & 0.729 & 0.844 & \textbf{0.925} & 0.720 & 0.867 & \textbf{0.919}\\
USPS & \textbf{1.000} & 0.998 & 0.997 & 0.975 & \textbf{0.979} & 0.978 & 0.682 & \textbf{0.923} & 0.912 & 0.960 & \textbf{0.998} & 0.988 & 0.851 & 0.857 & \textbf{0.934} & 0.813 & 0.894 & \textbf{0.940}\\
Weather & \textbf{1.000} & 0.988 & 0.986 & 0.934 & \textbf{0.945} & 0.942 & 0.662 & \textbf{0.939} & 0.852 & 0.931 & \textbf{0.998} & 0.957 & 0.739 & 0.856 & \textbf{0.927} & 0.766 & 0.860 & \textbf{0.912}\\
Wifi & \textbf{1.000} & 0.991 & 0.989 & 0.937 & \textbf{0.949} & 0.947 & 0.645 & \textbf{0.946} & 0.916 & 0.933 & \textbf{0.999} & 0.982 & 0.728 & 0.867 & \textbf{0.948} & 0.761 & 0.891 & \textbf{0.946}\\
OoD-Animals & \textbf{1.000} & 0.998 & 0.995 & 0.954 & \textbf{0.960} & 0.959 & 0.489 & \textbf{0.945} & 0.914 & 0.934 & \textbf{0.999} & 0.983 & 0.672 & 0.869 & \textbf{0.943} & 0.718 & 0.902 & \textbf{0.946}\\

\midrule
Average & \textbf{1.000} & 0.995 & 0.993 & 0.962 & \textbf{0.970} & 0.969 & 0.591 & \textbf{0.926} & 0.893 & 0.919 & \textbf{0.998} & 0.978 & 0.775 & 0.858 & \textbf{0.933} & 0.766 & 0.885 & \textbf{0.932} \\
\bottomrule
\end{tabular}
\label{table:full_40}
\end{table*}

\clearpage
\bibliographystyle{abbrv-doi-hyperref}
\bibliography{appendix-ref-part}